%% file: TaxonomyPreprint.tex
\documentclass[12pt,indent]{article}%a5paper
\usepackage{a4,latexsym,amsmath,amssymb}
\usepackage[latin1]{inputenc}
\usepackage{float}

\usepackage{booktabs,caption}
\usepackage[flushleft]{threeparttable}
\usepackage{rotating}

\usepackage{natbib}
\usepackage{graphics}
\usepackage[english]{babel}
\usepackage{tabularx}
\usepackage{lscape}
\usepackage{multirow}
\usepackage{rotating}
\usepackage{dsfont}
\usepackage{caption}
\usepackage{subcaption}
\usepackage{bbm}
\usepackage[table]{xcolor}
\usepackage{booktabs}
\usepackage{calc}
\usepackage{ifthen}
\usepackage{url}
\usepackage{colortbl}      %color the rows/columns of table using  >{\columncolor[gray]{0.8}}
\usepackage{tikz}

\usepackage{blindtext}
\usepackage{scrextend}
\usepackage{mathtools}
\usepackage{verbatim}

\usetikzlibrary{shapes,snakes,matrix}

\newenvironment{tabularsmall}
{ \footnotesize \sffamily \tabular } {
\endtabular
\normalfont }
\input{def}

\usepackage{natbib}

\usepackage{geometry}
\usepackage{setspace}
\usepackage{tikz}
\usepackage[]{graphicx}
\usepackage[]{color}
\makeatletter
\def\maxwidth{ %
  \ifdim\Gin@nat@width>\linewidth
    \linewidth
  \else
    \Gin@nat@width
  \fi
}
\makeatother

\definecolor{fgcolor}{rgb}{0.345, 0.345, 0.345}

\usepackage{framed}
\makeatletter
 {\par\unskip\endMakeFramed%
 \at@end@of@kframe}
\makeatother

\definecolor{shadecolor}{rgb}{.97, .97, .97}
\definecolor{messagecolor}{rgb}{0, 0, 0}
\definecolor{warningcolor}{rgb}{1, 0, 1}
\definecolor{errorcolor}{rgb}{1, 0, 0}
\usepackage{alltt}
\IfFileExists{upquote.sty}{\usepackage{upquote}}{}

\begin{document}
\bibliographystyle{chicago}
\sloppy
%%%%%%%%%%%%%%%%%%%%%%%%%%%%%%%%%%%%%%%%%%%%%%%%%%%%%%%%%%%%%%%%%%%%%
%                                                                   %
%         Definition einer modifizierten Kapitel�berschrift         %
%                                                                   %

\makeatletter
\renewcommand{\section}{\@startsection{section}{1}{\z@}%
        {-3.5ex \@plus -1ex \@minus -.2ex}%
        {1.5ex \@plus.2ex}%
        %{\reset@font\Large\sffamily}}
        {\reset@font\large\sffamily}}%{\bfseries bold}
\renewcommand{\subsection}{\@startsection{subsection}{1}{\z@}%
        {-3.25ex \@plus -1ex \@minus -.2ex}%
        {1.1ex \@plus.2ex}%
        %{\reset@font\large\sffamily\flushleft}}
        {\reset@font\normalsize\sffamily\flushleft}}
\renewcommand{\subsubsection}{\@startsection{subsubsection}{1}{\z@}%
        {-3.25ex \@plus -1ex \@minus -.2ex}%
        {1.1ex \@plus.2ex}%
        {\reset@font\normalsize\sffamily\flushleft}}
\makeatother

%                                                                   %
%%%%%%%%%%%%%%%%%%%%%%%%%%%%%%%%%%%%%%%%%%%%%%%%%%%%%%%%%%%%%%%%%%%%%

%%%%%%%%%%%%%%%%%%%%%%%%%%%%%%%%%%%%%%%%%%%%%%%%%%%%%%%%%%%%%%%%%%%%%
%                                                                   %
%         Definition einer modifizierten Bildunterschrift           %
%                                                                   %

\newsavebox{\tempbox}
\newlength{\linelength}
\setlength{\linelength}{\linewidth-10mm} \makeatletter
\renewcommand{\@makecaption}[2]
{
  \renewcommand{\baselinestretch}{1.1} \normalsize\small
  \vspace{5mm}
  \sbox{\tempbox}{#1: #2}
  \ifthenelse{\lengthtest{\wd\tempbox>\linelength}}
  {\noindent\hspace*{4mm}\parbox{\linewidth-10mm}{\sc#1: \sl#2\par}}
  {\begin{center}\sc#1: \sl#2\par\end{center}}
}

%                                                                   %
%%%%%%%%%%%%%%%%%%%%%%%%%%%%%%%%%%%%%%%%%%%%%%%%%%%%%%%%%%%%%%%%%%%%%

%\bibliographystyle{chicago}
%\baselineskip7mm
%\parindent 0.5cm
%\parskip2ex plus0.5ex minus 0.5ex
%\setlength{\parskip}{7pt plus 1pt minus 1pt}

\def\R{\mathchoice{ \hbox{${\rm I}\!{\rm R}$} }
                   { \hbox{${\rm I}\!{\rm R}$} }
                   { \hbox{$ \scriptstyle  {\rm I}\!{\rm R}$} }
                   { \hbox{$ \scriptscriptstyle  {\rm I}\!{\rm R}$} }  }

\def\N{\mathchoice{ \hbox{${\rm I}\!{\rm N}$} }
                   { \hbox{${\rm I}\!{\rm N}$} }
                   { \hbox{$ \scriptstyle  {\rm I}\!{\rm N}$} }
                   { \hbox{$ \scriptscriptstyle  {\rm I}\!{\rm N}$} }  }

\def\d{\displaystyle}\def\d{\displaystyle}

%\title{Binary Item Response Models as Building Blocks of Polytomous Latent Trait Models\\- a Common Framework for Classical  and Tree-Based Models}%  and the 
\title{A Taxonomy of Polytomous Item Response Models }%  and the 
%Ordering of Parameters in the Partial Credit Model }
  \author{Gerhard Tutz \\{\small Ludwig-Maximilians-Universit\"{a}t M\"{u}nchen}\\{\small Akademiestra{\ss}e 1, 80799 M\"{u}nchen}}
%\author{Gerhard Tutz \\{\small Ludwig-Maximilians-Universit\"{a}t M\"{u}nchen}\\{\small Akademiestra{\ss}e 1, 80799 M\"{u}nchen} \\ \large {Clemens Draxler} \\ \small {UMIT -- The Health and Life Sciences University} \\ \small {Eduard-Walln\"ofer-Zentrum 1, 6060 Hall in Tirol}}

%\author{Jan Gertheiss\footnote{To whom correspondence should be
%addressed: \texttt{jan.gertheiss@stat.uni-muenchen.de.}}
%\footnote{Department of Statistics, Ludwig-Maximilians-Universit�t
%Munich, Germany.} \ \& Gerhard TutY\footnotemark[2]}

% \ead{tutY@stat.uni-muenchen.de}
% \address{Ludwig-Maximilians-Universit\"{a}t M\"{u}nchen, Ludwigstra{\ss}e 33, D-80539 M\"{u}nchen, Germany}
%\author{LorenY Uhlmann}
%\ead{tutY@stat.uni-muenchen.de}
%\address[muc]{Ludwig-Maximilians-Universit\"{a}t M\"{u}nchen, Akademiestra{\ss}e 1, 80799 M\"{u}nchen, Germany}
%\cortext[cor]{Corresponding author. Tel.: ++4989 2180 3044; fax.:
%++4989 2180 5308.}
%{\texttt{\small \{tutY, uhlmann\}@stat.uni-muenchen.de}}}
%\address[muc1]{Ludwig-Maximilians-University Munich, Ludwigstrasse 33, D-80539 Munich, Germany}
%\address[muc2]{Ludwig-Maximilians-University Munich, Akademiestra{\ss}e 1, D-80799 Munich, Germany}
%\cortext[cor]{Corresponding author. Tel.: ++49 89 2180 3044; fax.:
%++49 89 2180 ???.}
\maketitle
 %\doublespacing
\begin{abstract} % \renewcommand{\baselinestretch}{1.3} \small\normalsiYe
\noindent
A common framework is provided that comprises classical ordinal item response models as the cumulative, sequential and adjacent categories models as well as   nominal response models and item response tree models. The   taxonomy  is based on the   ways  binary models can be seen as building blocks of  the various models. In particular one can distinguish between conditional and unconditional model components. Conditional models are by far the larger class of models containing the adjacent categories model  and the whole class of hierarchically structured models. The latter is introduced as a class of models that comprises binary trees and hierarchically structured models that use ordinal models conditionally. 
The study of the binary models contained in  latent trait models  clarifies the relation between models and the interpretation of  item parameters. It is also used to distinguish between ordinal and nominal models by giving a conceptualization of ordinal models. The taxonomy differs from previous  taxonomies by focusing  on the structured use of dichotomizations instead of the role of parameterizations.  
%The taxonomy for ordinal models also contains a new general class of hierarchically structured models, which can be seen as a generalization of item response tree
%models. For this class of models estimation methods are developed, which make use of commonly available program packages.
%Ordered item response models that are in common use  can be divided into three groups, cumulative, sequential and adjacent categories model.The models
\end{abstract}


%\doublespacing
\section{Introduction}

%Over the past decades various latent trait models for ordered response data have been proposed in the literature. Widely used models are in particular the graded response model \citet{samejima1995acceleration,samejima2016graded}, the so-called partial credit model \citet{Masters:82} and  the sequential or step model \citet{Tutz:90b, verhelst1997steps}. A more recently  introduced class of models are item response trees, which allow new ways to model ordered responses, see, for example, \citet{de2012irtrees, bockenholt2013modeling}. An overview of classical models including nominal ones has been given by \citet{VanderLind2016}.  

Various latent trait models for ordered response data have been proposed in the literature, for an overview see, for example, \citet{VanderLind2016}.  One can in particular distinguish between three basic types of models, cumulative models, sequential models and adjacent categories models. One of the objectives of the present paper is to show how these models  are easily built from binary latent trait models. The way how the binary models are used to construct models helps to understand the structure of the models and to clarify the meaning of the parameters. It also provides a framework that allows to embed  more recently developed ordinal item response models as, for example,  tree-based models,  yielding a general taxonomy of ordinal item response models.

The proposed taxonomy is quite different from that given by \citet{thissen1986taxonomy}. Their classification into ``difference'' models and ``divide-by-total'' models   is based on the form of the response probabilities, which may be represented as differences or as a ratio obtained by dividing by sums of terms. The third class of models they consider are ``left-side added models'', which arise if the parameterization is extended to account for guessing parameters.
The taxonomy proposed here is of a  different nature. It is based on exploiting how    ordinal models can be constructed by using  (conditional or unconditional) dichotomizations of response categories. It also works the other way, by clarifying which binary models (or dichotomizations) are contained in  ordinal models. By investigating this structural aspect one obtains a taxonomy that uses that ordinal models can be characterized by the way they determine the choice of specific subsets of categories. 

As a preview let us consider how classical models can be described by considering dichotomizations. The basic building blocks are binary models, which  
in its simplest form specify that the responses of   person $p$ on item $i$ are determined by  
\begin{equation}\label{eq:bin}
P(Y_{pi}=1)=F(\theta_p-\delta_{i}),
\end{equation} 
where $F(.)$ is a cumulative distribution function, $\theta_p$ is the person parameter, and $\delta_{i}$ is the item parameter, typically a difficulty or threshold.  An important member of this class of models is the Rasch model, which is obtained if $F(.)$ is the logistic distribution function $F(\eta)=\exp(\eta)/(1+\exp(\eta))$.

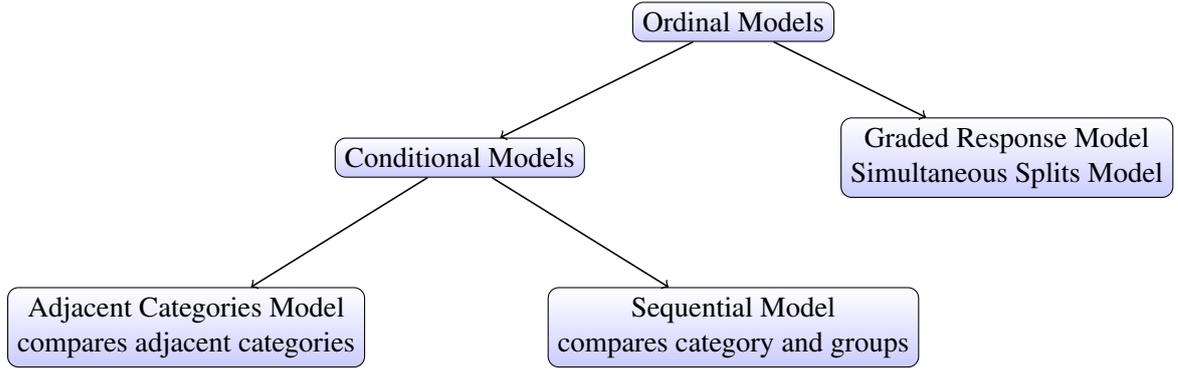
\begin{figure}
\begin{center}
\begin{tikzpicture}[scale=.9, transform shape]

\begin{scope}[every node/.style={shape=rectangle, rounded corners,
    draw, align=center,top color=white, bottom color=blue!20}]
\node[xshift=0 cm,yshift=0cm] (c0){Conditional Models};
\node[xshift=-4 cm,yshift=-2.5cm] (c1){ Adjacent Categories Model\\compares adjacent categories};
 \node[xshift=4 cm,yshift=-2.5cm] (c2){ Sequential Model\\compares category and groups};
 %\node[xshift=-1cm,yshift=-5cm] (c3){ Sequential Model\\compares category and groups};
  %\node[xshift=4cm,yshift=-5cm] (c4){ IRTrees\\conditional binary splits};
 %\node[xshift=9cm,yshift=-5cm] (c6){ Hierarchical Trees\\conditional ordinal models};
\node[xshift=8cm,yshift=0cm] (c5){Graded Response Model\\Simultaneous Splits Model};
  \node[xshift=4cm,yshift=2cm] (c00){Ordinal Models}; 
 \end{scope}

\draw[->, semithick] (c0) -- (c1);
\draw[->, semithick] (c0) -- (c2);
%\draw[->, semithick] (c2) -- (c3);
 %\draw[->, semithick] (c2) -- (c4);
 %\draw[->, semithick] (c2) -- (c6);
\draw[->, semithick] (c00) -- (c0);
 \draw[->, semithick] (c00) -- (c5); 
%\node[xshift=8cm,yshift=-0.50cm] (c2){Query Agreement/Disagreement};
%\node[xshift=8cm,yshift=-2cm] (c2){Query Extremity};
%\node[xshift=8cm,yshift=-4cm] (c2){Query Weakness of Attitude};
\end{tikzpicture}
 \caption{Structure of classical ordinal latent trait models.}
\label{struc1}
\end{center}
\end{figure}

Given one has a response in \textit{ordered} categories $\{0,1,\dots,k\}$ there are several ways  to construct an ordinal model from binary models of the form (\ref{eq:bin}). The binary models
can be used to compare specific categories or groups of categories from $\{0,1,\dots,k\}$. One can, in particular, 

\begin{itemize}
\item[-] compare groups of categories that result from splitting the categories into the subsets $\{0,1,\dots,r-1\}$ and $\{r,\dots,k\}$, 
\item[-] compare (conditionally) between two categories, for example, adjacent categories,
\item[-] compare (conditionally) between a category and a set of adjacent categories, for example,  $\{r-1\}$ and $\{r,\dots,k\}$.
\end{itemize}
The different ways to compare categories correspond to  cumulative models, adjacent categories and sequential models in that order.  The taxonomy given in  Figure \ref{struc1} distinguishes  between conditional and non-conditional models, a distinction which follows from the consideration of the binary models that are contained in the ordinal models. The cumulative or graded response model is the only non-conditional model from this class of models. The other two use some sort of conditioning in the binary building blocks within the models.

The graded response model corresponds to the difference models, and the adjacent categories models to the divide-by-total models in the Thissen-Steinberg taxonomy. \citet{thissen1986taxonomy} did not consider sequential models, which were not known in 1986. Although in the present taxonomy and the Thissen-Steinberg taxonomy   two model classes cover the same types of models, the focus is different. Here, the models are not characterized by the form of the probability but by their  building blocks. The consideration of building blocks allows not only to include the sequential models and other models, but is also helpful to obtain a valid interpretation of the parameters of the models, which has not always been correct  and has been a subject of debate in  the literature on ordinal models, see, for example, \citet{adams2012rasch}, \citet{andrich2013expanded,andrich2015problem}, \citet{garcia2017analysis}, \citet{Tu2020JMP}.   

The focus on building blocks allows to identify common structures beyond the choice of the response function $F(.)$ in (\ref{eq:bin}) and the parameterization. For the structuring it does not matter if one chooses the logistic response function, the normal ogive or any other strictly monotone distribution function. One might also extend the linear term 
to include an item discrimination parameter by using  $\alpha_i(\theta_p-\delta_{i})$ instead of $\theta_p-\delta_{i}$, or extend it to include  guessing parameters.   Although parameterization is considered secondary when characterizing model types it can be important as a source of ordinality in latent trait models. This will be investigated separately when considering nominal models.

In Section \ref{sec:Class} classical ordered response models are considered. It is used that they can be characterized as parameterizations of split variables, which yield the preliminary structure given in Figure \ref{struc1}.  Section \ref{sec:HierStr}   is devoted to tree structured models. In particular binary IR-tree models are considered, which have been introduced as flexible models to account for response styles. Within the general hierarchically structured models it is distinguished between binary IR-tree models and partitioning models that contain ordinal building blocks. Both  are embedded into the proposed framework yielding the  taxonomy given in Section \ref{sec:tax}.  In Section \ref{sec:NomPar}  the role of parameterizations and the order in ordinal models are discussed. It is in particular investigated how ordinal models can be obtained by using constraints on parameters in nominal models. 
The final chapter completes the taxonomy by including the  class of finite mixture models, which are divided into homogeneous and heterogeneous mixture models.

\section{Classical Ordered Response Models}\label{sec:Class}

In the following let $Y_{pi} \in \{0,1,\dots,k\}$, $p=1,\dots,P$, $i=1,\dots,I$, denote the ordinal response of person $p$ on item $i$.
An important partition of the response categories is the partition into the subsets $\{0,\dots,r-1\}$ and $\{r,\dots,k\}$, which can be represented by  
the binary variable 
\begin{equation}\label{eq:cond2}
Y_{pi}^{(r)}=\left\{
\begin{array}{ll}
1&Y_{pi}  \ge r \\
0&Y_{pi} < r.
\end{array} 
\right.
\end{equation}
The variables $Y_{pi}^{(1)},\dots,Y_{pi}^{(k)}$ are called \textit{split variables} because they split the response categories into two subsets. As shown by \citet{Tu2020JMP} they play a major role in the construction of the traditional ordered latent trait models. In the following we use these results to derive a taxonomy of the traditional models

\subsection{Simultaneous Modelling of Splits: The Graded Response Model}\label{sec:Cum}

Let us assume that the response categories represent levels of performance in an achievement test.  Then one can consider two groups of categories,
$\{0,1,\dots,r-1\}$ for low performance and $\{r,\dots,k\}$  for high performance, where low and high are relative terms that refer to ``below category $r$''
and ``above or in category r''. One might assume that the split into low and high performance is determined by a binary  model with person ability $\theta_p$
and a threshold that depends on the category at which the categories have been split by specifying 
\begin{equation}\label{eq:cumbin}
P(Y_{pi}^{(r)}=1)=F(\theta_p-\delta_{ir},)\quad, r=1,\dots,k.
\end{equation} 
Thus, for each dichotomization into categories $\{0,1,\dots,r-1\}$ and $\{r,\dots,k\}$ a binary model is assumed to hold. 
Importantly, the models are assumed to hold \textit{simultaneously} with the same person ability $\theta_p$ but different item difficulties $\delta_{ir}$. 
Simple rewriting yields the \textit{cumulative model} 
\begin{equation}\label{eq:cum}
P(Y_{pi} \ge r)=F(\theta_p-\delta_{ir}),\quad  r=1,\dots,k,
\end{equation}
which is equivalent to a version of Samejima's \textit{graded response model} \citep{samejima1995acceleration,samejima2016graded}.  %Samejima introduced it initially as a normal-ogive model.
Thus, the graded response model can be seen as a model for which the dichotomizations into the categories $Y_{pi} < r$ and $Y_{pi} \ge r$ are \textit{simultaneously} modeled. One consequence is that item difficulties are ordered. 
Since $P(Y_{pi} = r)=P(Y_{pi} \ge r)-P(Y_{pi} \ge r+1)=F(\theta_p-\delta_{ir})-F(\theta_p-\delta_{i,r+1}) \ge 0$, one obtains that  $\delta_{ir} \le\delta_{i,r+1}$has to hold  for all categories. 

The strong link between the binary responses and the ordinal response yields a specific view of the graded response model that differs from traditional ones. 
In an achievement test the sequence of binary responses $ (Y_{pi}^{(1)},\dots, Y_{pi}^{(k)})$ can be seen as referring  to tasks with increasing difficulties. More concrete, because item difficulties are ordered, one has $P(Y_{pi}^{(r)}=1) \ge P(Y_{pi}^{(r+1)}=1)$, which means the ``task'' represented by $Y_{pi}^{(r)}$ is simpler than the ``task'' $Y_{pi}^{(r+1)}$. Moreover, if the task   $Y_{pi}^{(r)}$ was completed ($Y_{pi}^{(r)}=1$ or, equivalently, $Y_{pi} \ge r$), the simpler tasks $Y_{pi}^{(s)}, s < r$ ($Y_{pi}^{(s)}=1$ or, equivalently, $Y_{pi} \ge s$) were also completed. Therefore, the outcome of the sequence of binary variables  has the specific form 
\[
(Y_{pi}^{(1)}\dots,Y_{pi}^{(k)})=(1,\dots,1,0,\dots,0),
\]
which means a sequence of ones is followed by a sequence of zeros. Binary variables that follow  this pattern have been called \textit{Guttman variables} and the resulting response space is usually referred to as Guttman space, a term that was used by \citet{andrich2013expanded} when discussing partial credit models.  
%The increasing sequence of difficulties $\delta_{i1} \le \dots \le\delta_{ik}$ may also be seen as thresholds that have to be exceeded to obtain a higher level of performance.  $Y_{pi}^{(r)}=1$ means that threshold $\delta_{ir}$ has been exceeded. 
%%%%%%%%%%%%%%%%%
\blanco{
A nice feature is that the ordinal response is simply given as the sum of the binary variables
\[
Y_{pi} = Y_{pi}^{(1)}+ \dots + Y_{pi}^{(k)},
\]
which means the ordinal response is  the number of tasks that have been successfully completed.
}
%%%%%%%%%%%%%%%%%%%%%%%%

The more classic derivation of the cumulative model suggests that the item parameters may be seen as thresholds. 
Let  $\tilde Y_{pi}=\theta_p+\varepsilon_{pi}$, where   $\varepsilon_{pi}$ is a noise variable with symmetric continuous distribution function $F(.)$, denote a   latent variable that is invoked if person $p$ tries to solve item $i$.  $\tilde Y_{pi}$ is essentially the ability of the person plus a noise variable and can be seen as the random ability of the person.  The category boundaries approach assumes that   category $r$ is observed if the latent variable is between thresholds $\delta_{ir}$ and $\delta_{i,r+1}$. More formally, one has $Y_{pi}=r\quad\Leftrightarrow\quad\delta_{ir}\leq \tilde Y_{pi}<\delta_{i,r+1}$. It is easily seen that one obtains the cumulative model and thresholds have to be ordered.

%%%%%%%%%%%%%%%%%%%%%%%%%%%%%%%

\citet{thissen1986taxonomy} called the graded response models  ``difference'' models because the probabilities are given as differences, $P(Y_{pi} = r)=F(\theta_p-\delta_{ir})-F(\theta_p-\delta_{i,r+1})$. Although they also start with binary models they do not further investigate that the models have to hold simultaneously.

\subsection{Conditional Comparison of Categories: the  Partial Credit and   General Adjacent Categories Models}\label{sec:Cum}

Rather than compare groups of categories by utilizing a binary model one can also compare two categories from the set of categories $\{0,1,\dots,k\}$. 
A choice that suggests itself are adjacent categories. Let the binary models that compare  two adjacent categories be given by 
\begin{equation}\label{eq:part}
P(Y_{pi} = r | Y_{pi} \in \{r-1,r\}) = F(\theta_p-\delta_{ir}),\quad r=1,\dots,k.
\end{equation}
Again all the models contain the same person parameter but model-specific item parameters. For the logistic distribution function one obtains the \textit{partial credit model}  
\[
P(Y_{pi}=r)= \frac{\exp(\sum_{l=1}^{r}(\theta_p-\delta_{il}))}{\sum_{s=0}^{k}\exp(\sum_{l=1}^{s}(\theta_p-\delta_{il}))}, \quad r=1,\dots,k,
\]
which was propagated by  \citet{Masters:82} and \citet{MasWri:84}. It is  equivalent to the \textit{polytomous Rasch model}, which is just a different parameterization, see, for example, \citet{andrich2010sufficiency}.  \citet{thissen1986taxonomy} called the partial credit model  model a  ``divide-by-total'' model because of the denominator in the probabilities.  However, the family of adjacent categories models is much larger because in (\ref{eq:part}) any strictly monotone distribution function can be used, for example, the use of the normal distribution yields a probit version of the adjacent categories model. In the logistic version  sufficient statistics for item and person parameters are available. While the existence of sufficient statistics is an advantage if one wants to estimate parameters conditionally  it is of lesser importance if one uses marginal estimates. 
%Then goodness-of-fit might be more important than sufficient statistics. 
We refer to this model class more generally as \textit{adjacent categories models}, of which the polytomous Rasch model or partial credit model are just the most prominent members.

%If one uses binary models as building blocks, the question arises why one should confine oneself to adjacent categories, although they seem a natural choice. An alternative would be to use binary models for a set of pairs of categories $(s_1,r_1),\dots,(s_k,r_k)$, $s_i<r_i$. It can be shown that postulating  binary models for pairs of categories also  yields the partial credit model. Therefore, the partial credit model can be seen as a general model for pairs of categories.

An alternative form of the partial credit model, which  emphasizes the implicit comparison of categories is
\begin{equation}\label{eq:adj2}
\log \left(\frac {P(Y_{pi}=r)}{P(Y_{pi}=r-1)}\right)=  \theta_p-\delta_{ir},\quad r=1,\dots,k.
\end{equation}
That means, the PCM directly compares two adjacent categories, and $\theta_p$ determines the strength of the preference for the higher category.

It should be emphasized that the binary models used as building blocks  are \textit{conditional} models, it is assumed that a binary model holds \textit{given the response is in two categories from the set of available categories}. This is seen from the representation (\ref{eq:part}) but hidden in the representation (\ref{eq:adj2}). However, it has consequences for the interpretation of parameters. The item parameters represent thresholds \textit{given} the response is in categories $\{r-1,r\}$ and the trait parameters are the abilities to score $r$ rather than $r-1$ \textit{given} the response is in categories $\{r-1,r\}$. 
Therefore, the parameters refer to a local conditional decision or preference although changing the item parameter changes the probabilities of all possible outcome values since the PCM assumes that the binary models hold simultaneously. 

%Nevertheless the binary models are conditional models and parameters should be interpreted with reference to the conditional structure. One consequence of the conditional parameterization is that thresholds do not have to be ordered though there has been some discussion on the ordering of thresholds \citep{adams2012rasch,andrich2013expanded,andrich2015problem}. 

The conditional structure is also seen if the model is represented by using split variables. Since the condition $Y_{pi}^{(r-1)}=1,Y_{pi}^{(r+1)}=0$ is equivalent to $Y_{pi} \in \{r-1,r\}$ one obtains that  the PCM is equivalent to postulating for all split-variables
\[
P(Y_{pi}^{(r)}=1|Y_{pi}^{(r-1)}=1,Y_{pi}^{(r+1)}=0)=  F(\theta_p-\delta_{ir}),
\]
where $F(.)$ is the logistic distribution function. It means that a Rasch model holds for the split-variable $Y_{pi}^{(r)}$ given the  split $Y_{pi}^{(r-1)}$ is in favor of higher categories while the split $Y_{pi}^{(r+1)}$ is in favor of lower categories.

The class of adjacent categories model also contains simplified versions that use  sparser parameterizations. By assuming that the item parameters can be decomposed into two terms in the form $\delta_{il}= \delta_i+\tau_{l}$, one obtains the  \textit{Rasch rating scale model}   \citep{andrich1978rating, Andrichh2016}. The model can also be extended to include a slope parameter if it is included in the binary model that distinguishes between adjacent categories \citep{muraki1990fitting, muraki1997generalized}.

\subsection{Conditional Comparison of a Single Category and a Group of Categories:   Sequential Models}

In achievement tests  frequently items are used that are solved in  consecutive observed steps. For example, a  mathematical problem may have the form:  $(\sqrt{49}-9)^3=?$. One can distinguish four levels: no problem solved (level 0), $\sqrt{49}=7$ solved (level 1), $7-9=-2$ solved (level 2), $(-2)^3=-8$ solved (level 3). Obviously the sub problems have to be solved in a consecutive way. A sub problem can only be solved if the all the previous sub problems have been solved.
A model that explicitly models the solving of sub problems has the form 
%Let us consider the subset of categories $\{r-1,\dots,k\}$, which is partitioned into the groups $\{r-1\}$ and $\{r,\dots,k\}$. Let a conditional binary model determine the response in these groups by postulating  
%\begin{equation}\label{eq:sequ}
%P(Y_{pi} \in \{r,\dots,k\} | Y_{pi} \in \{r-1,\dots,k\}) = F(\theta_p-\delta_{ir}),\quad r=1,\dots,k.
%\end{equation}
%or equivalently 
\begin{equation}\label{eq:sequ2}
P(Y_{pi} \ge r | Y_{pi} \ge r-1) = F(\theta_p-\delta_{ir}),\quad r=1,\dots,k.
\end{equation}
The model is known as \textit{sequential model} \citep{Tutz:90b} or \textit{step model} \citep{verhelst1997steps}. 
It is  a process model for consecutive steps. One models the transition to higher categories given the previous step was successful. The first step is the only non-conditional step. If it fails, the response is in category $0$ (first sub problem not solved), if it is successful, the response is larger than 0 (first sub problem solved).  In the latter case the   person tries to take the second step. If it is not successful, the response is in category $1$ (second sub problem not solved), if it is successful, the response is larger than 1 (second sub problem solved), etc. In the $r$-th step it is distinguished between $Y_{pi} = r-1$ and $Y_{pi} \ge r$   \textit{given} at least level $r-1$ is reached ($Y_{pi} \ge r-1$). In the model the parameter $\theta_p$ represents the person's ability to successfully perform each of the steps while $\delta_{ir}$ is the difficulty in step $r$. Of course, later steps can be easier than early steps, thus item difficulties are not necessarily ordered. In the example step 2 ($7-9$) is certainly easier to master than step 1 ($\sqrt{49}=7$). However, sub problem 2 can be only solved after step 1 was successful. Therefore, the item parameters have \textit{local meaning}, they refer to the difficulty in a step given that all previous steps were successful. In contrast, the same ability parameter is present in each of the steps, which makes the model uni-dimensional in terms of person parameters.

\begin{figure}
\begin{center}
\begin{tikzpicture}[scale=.9, transform shape]
\begin{scope}[every node/.style={shape=ellipse,
    draw, align=center,top color=white, bottom color=blue!20}]
\node (c0){$0,1,2,3$};
\node[xshift=-2cm,yshift=-2cm] (c1){  $0$};
 \node[xshift=2cm,yshift=-2cm] (c2){  $1,2,3$};
\node[xshift=0cm,yshift=-4cm] (c3){ $1$};
\node[xshift=4cm,yshift=-4cm] (c4){ $2,3$};
\node[xshift=2cm,yshift=-6cm] (c5){ $2$};
\node[xshift=6cm,yshift=-6cm] (c6){ $3$};
\end{scope}
\draw[->, semithick] (c0) -- (c1);
\draw[->, semithick] (c0) -- (c2);
 \draw[->, semithick] (c2) -- (c3);
 \draw[->, semithick] (c2) -- (c4);
 \draw[->, semithick] (c4) -- (c5);
\draw[->, semithick] (c4) -- (c6);
%\node[xshift=8cm,yshift=0cm] (c6){ Root};
\node[xshift=8cm,yshift=-1cm] (c6){ First Step};
\node[xshift=9cm,yshift=-2cm] (c7){ {\small Unsuccessful: level 0}};
\node[xshift=8cm,yshift=-3cm] (c6){ Second Step};
\node[xshift=9cm,yshift=-4cm] (c7){ {\small Unsuccessful: level 1}};
\node[xshift=8cm,yshift=-5cm] (c6){ Third Step};
\node[xshift=9cm,yshift=-6cm] (c7){ {\small Unsuccessful: level 2}};
\end{tikzpicture}
\caption{The sequential model as a hierarchically structured model.}
\label{Treseq}
\end{center}
\end{figure}
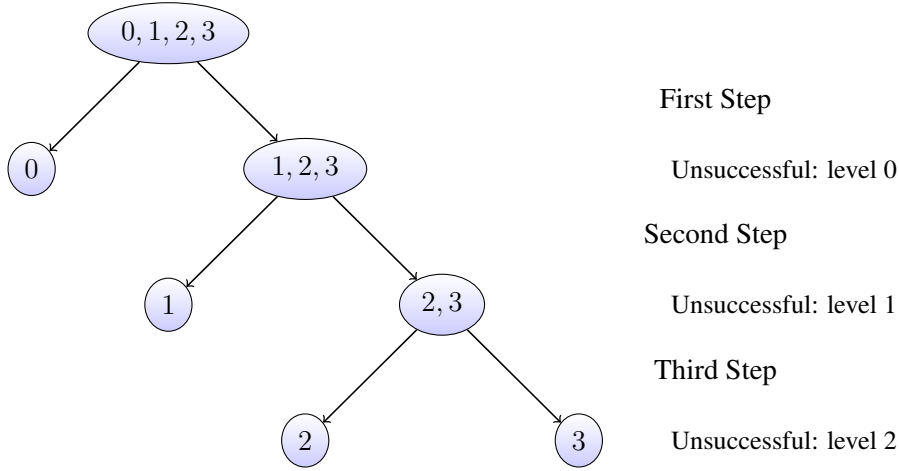

The logistic version of the model, also called logistic sequential model, can be given in the alternative form of a  \textit{continuation ratio model},
\begin{equation}\label{eq:sequ3}
\log \left(\frac {P(Y_{pi}\ge r)}{P(Y_{pi}=r-1)}\right)=  \theta_p-\delta_{ir},\quad r=1,\dots,k,
\end{equation}
\citep{Agresti:2013}. The logits  on the left hand side compare the categories the probability of a response in the categories $\{r,\dots,k\}$ to the probability of a response in category $\{r-1\}$. In this sense the binary models contained in the sequential model compare  groups of categories to   single categories. 
This comparison is also seen from the tree representation of the model given in Figure \ref{Treseq}, which shows the sequence of (conditional) binary splits in a sequential model with four categories.  In the  $r$-th step a decision between category $\{r-1\}$ and categories $\{r,\dots,k\}$ is obtained. The split is conditional, given categories $\{r-1,\dots,k\}$, that means, under the condition that the previous step was successful.

A disadvantage of the model representation (\ref{eq:sequ3}) is that it does not directly show the underlying process. The implicit conditioning on responses $Y_{pi}\ge r$, which is essential for the interpretation of the model parameters, gets lost. It is however seen in the model representation with \textit{split variables} given by

\begin{equation}\label{eq:cumbin3}
P(Y_{pi}^{(r)}=1|Y_{pi}^{(r-1)}=1,\dots,Y_{pi}^{(1)}=1)=F(\theta_p-\delta_{ir})\quad, r=1,\dots,k,
\end{equation}

which again shows that the split variables $(Y_{pi}^{(1)}\dots,Y_{pi}^{(k)})$ form a Guttman space.

A rating scale version of the model, in which the parameter $\delta_{ir}$ is split up into a an item location parameter $\delta_{i}$ and a step
parameter $\tau_r$, with $\sum_r \tau_r=0$ has been considered by \citet{Tutz:90b}, extended versions with predictor $\alpha_{ir}\theta_p-\delta_{ir}$ and nonparameteric versions have been considered by \citet{hemker2001measurement}.

\begin{table}[h!]
\centering
\caption{Overview of traditional ordinal models.}\label{tab:over}
\begin{tabularsmall}{@{}lccc@{}}
\toprule \medskip
               & Category  Representation   &     Conditional  Representation      &  Conditional  Representation             \\
               &  Logistic Version&      General Version     &  With split variables                 \\\medskip
               &$\log (.)=  \theta_p-\delta_{ir}$       & $P(.)=F(\theta_p-\delta_{ir})$ & $P(.)=F(\theta_p-\delta_{ir})$\\
 \midrule \medskip
{\small Cumulative  }     &  $\log \left(\frac {P(Y_{pi} \ge r)}{P(Y_{pi}<r)}\right)$     &   $P(Y_{pi} \ge r)$               & $P(Y_{pi}^{(r)}=1)$                               \\
\medskip
{\small Adjacent  }  &    $\log \left(\frac {P(Y_{pi}=r)}{P(Y_{pi}=r-1)}\right)  $    &        $P(Y_{pi} = r | Y_{pi} \in \{r-1,r\}) $          &$P(Y_{pi}^{(r)}=1|Y_{pi}^{(r-1)}=1,Y_{pi}^{(r+1)}=0)$                                \\
\medskip
{\small Sequential  }  &    $\log \left(\frac {P(Y_{pi}\ge r)}{P(Y_{pi}=r-1)}\right)  $   &        $P(Y_{pi}\ge r | Y_{pi}\ge r-1) $          &$P(Y_{pi}^{(r)}=1|Y_{pi}^{(r-1)}=1)$                               \\

\bottomrule
\end{tabularsmall}
\end{table}

\subsection{Overview on Classical Ordinal Models }

The fact that all the models contain binary models that split categories into two subsets can be exploited to distinguish between models by focusing on the underlying conditioning. In particular,   
in the partial credit model and the sequential model the splits are conditional whereas in the cumulative model the splits are simultaneous  but not conditional. Figure \ref{struc1} visualizes the resulting hierarchy of models.

In Table \ref{tab:over} the models are given in various representations.
The left column shows the logistic versions of the models. It shows which categories or groups of categories are compared. In particular it is seen which type of logits are determined by the difference between person parameter and item parameter, $\theta_p-\delta_{ir}$. For example, in the partial credit model one has the adjacent categories logits $\log  (  {P(Y_{pi}=r)}/{P(Y_{pi}=r-1)} )$, in the sequential model one has the continuation ratios $\log ( {P(Y_{pi}\ge r)}/{P(Y_{pi}=r-1)})$. In the middle column the general conditional representations of the models are given. In these representations the distribution function $F(.)$ can be any strictly monotonic distribution function. It shows which \textit{conditional} binary response models are contained in the ordinal model. In the case of the graded response model the condition is empty since it is a non-conditional model. 

The right column shows the representation of the general models with split variables. It also shows clearly the conditioning implicitly contained in the models.   
It should be emphasized that in \textit{all} the models the split variables $(Y_{pi}^{(1)}\dots,Y_{pi}^{(k)})$ form a Guttman space with outcomes having the form $(1,\dots,1,0,\dots,0)$. They can be seen as generating the Guttman space, which is always defined and is not linked to any specific model, see \citet{Tu2020JMP}. 

%%%%%%%%%%%%%%%%%%
\blanco{
Moreover, the ordinal response is always given by $Y_{pi}=Y_{pi}^{(1)}+\dots+Y_{pi}^{(k)}$. These properties follow simply  from the definition of the split variables because one has
in general
\[
Y_{pi}=r \Leftrightarrow (Y_{pi}^{(1)}, \dots,Y_{pi}^{(r)},\dots, Y_{pi}^{(k)}) = (1,\dots,1,0,\dots,0),
\]
where $Y_{pi}^{(r)}$ is the last one of the sequence of split variables with a value 1. This general property makes the split variables a versatile tool that can be used to characterize the structure of models. Although it is well known that ordinal models contain binary models as given in the first two columns of Table \ref{tab:over} the link to the split variables seems not to have been exploited in a systematic way, however, it is important since it clarifies that all models specify dichotomizations of categories, conditionally or unconditionally. 
}
%%%%%%%%%%%%%%%%%%%%%%

The dichotomizations or Guttman variables also clarify the meaning of parameters. In the graded response model the split variables $Y_{pi}^{(r)}$, which distinguish between a response $Y_{pi} \ge r$
(strong performance) and $Y_{pi} < r$ (weak performance), are directly linked to the difference between 'ability' and item parameter, $\theta_p-\delta_{ir}$. 
The corresponding binary models for split variables are unconditional and have to hold simultaneously. This allows to see the item difficulties as thresholds that are necessarily ordered, and which  have to be exceeded to obtain higher levels. The variables $Y_{pi}^{(r)}$ should not be seen as steps. $Y_{pi}^{(r)}=1$ simply denotes that a person has at least performance level $r$. Since performance levels are ordered, that means, its performance cannot be below level $r$, or, in split variables, $Y_{pi}^{(1)}=\dots=Y_{pi}^{(r)}=1$, which is the Guttman property of the binary responses. One observes $Y_{pi}=r$, if, in addition $Y_{pi}^{(r+1)}=0$, which means that the performance is below level $r+1$.
However, no steps or transitions are needed to explain the level of performance. As \citet{andrich2015problem} argues, if a performance like acting is to be classified according to some protocol, the judge places the person's performance   in one of the categories on the trait, not how the person transitioned in getting to the category.  Even in simple binary models for problem solving one observes if the problem was solved or not, but not the transition. Thus, when considering ordinal models and the binary models contained in them there is no reason to construct a transition. It might be misleading and is not compatible with the underlying process, which is determined by \textit{simultaneous} dichotomizations or the placing on the continuum of the latent scale, which is divided by the thresholds  $\delta_{i1}\le \dots \le \delta_{i,k}$.

Interpretation of parameters is quite different in conditional models. Let us start with the sequential model since it is by construction  a step or transition model. The split variables representation, $P(Y_{pi}^{(r)}=1|Y_{pi}^{(r-1)}=1)=F(\theta_p-\delta_{ir})$, shows that the difference between ability and item difficulty determines if the performance is above or in category $r$ given   at  least performance level $r-1$ has been reached. It makes the parameter $\delta_{ir}$ a local threshold parameter. There is no ordering of thresholds involved since later steps might be easier than previous steps. 
%It should also be noted that the condition $Y_{pi}^{(r-1)}=1$ is equivalent to $Y_{pi}^{(1)}=\dots=Y_{pi}^{(r-1)}=1$ since the split variables form a Guttman space. The condition implies that all previous steps were successful.

In the partial credit model the decision that the performance is in category $r$ as a function of the difference between ability and item parameter
is under the condition $Y_{pi} \in \{r-1,r\}$, or equivalently, $Y_{pi}^{(r-1)}=1,Y_{pi}^{(r+1)}=0$. Thus, the 
binary models are conditional models and parameters should be interpreted with reference to the conditional structure. One consequence of the conditional parameterization is that thresholds do not have to be ordered though there has been some discussion on the ordering of thresholds \citep{adams2012rasch,andrich2013expanded,andrich2015problem,Tu2020JMP}.

Consideration of the binary models contained in ordinal models also explains why some models are  robust against the collapsing of categories. In general, if a model holds for the original categories it does not necessarily hold if adjacent categories are grouped yielding a smaller set of response categories, although that might be an attractive feature of a model, see also \citet{jansen1986latent} where the extreme case of dichotomization in  the polytomous Rasch model is considered in detail. The graded response model (with a logistic distribution function) holds also for dichotomized responses (or other groupings of adjacent categories) since the splits themselves follow a Rasch model. This is different for the adjacent categories and the sequential model. They are not robust against collapsing of categories  because the Rasch models that are contained are conditional. However, collapsing of categories changes the conditioning. If, for example, categories $r$and $r+1$ are collapsed to form a new category, the conditions in the binary submodels   after collapsing differ from the conditions in the original set of categories for all conditions that contain the new category.

\section{Hierarchically Structured Modeling:  Tree-Based Models}\label{sec:HierStr}
 
The classical models considered in the previous section represent different types of modelling concerning the conditioning. While the graded response model is a  model that does not rely on conditioning,   the partial credit model   conditions on a response in adjacent categories. The sequential model  is conditional but, in contrast to the partial credit model,    it can be represented as a tree (see Figure \ref{Treseq}). This makes it a special model, it is \textit{hierarchical}, that means,  it can be represented by a sequence of conditional splits.  Neither the graded response models nor the partial credit model are hierarchical.
More recently with  IRTrees a whole class of hierarchical models has been introduced. In the following we will first consider binary IRTrees and then consider alternative approaches.
For simplicity  in the following the response categories are $\{1, \dots,m\}$, which is the common notation in IR-Trees.

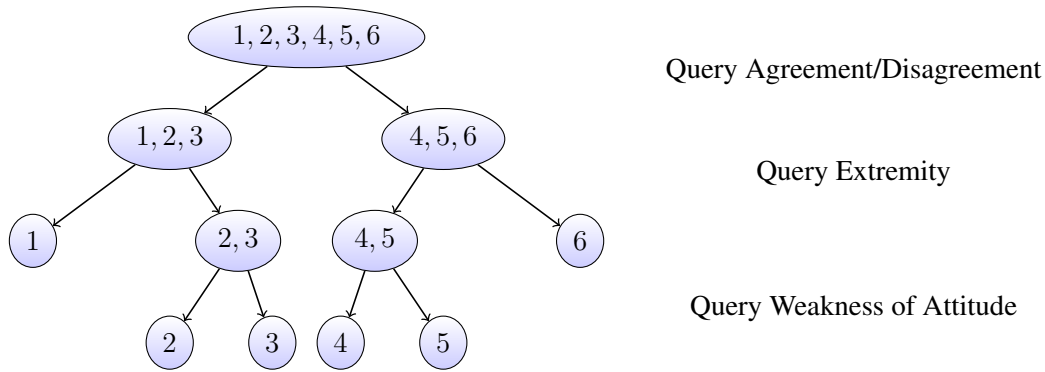
\begin{figure}
\begin{center}
\begin{tikzpicture}[scale=.9, transform shape]

%\begin{scope}[every node/.style={shape=rectangle, rounded corners,    draw, align=center,top color=white, bottom color=blue!20}]
\begin{scope}[every node/.style={shape=ellipse,    draw, align=center,top color=white, bottom color=blue!20}]
\node (c0){$1,2,3,4,5,6$};
\node[xshift=-2 cm,yshift=-1.5cm] (c1){ $1,2,3$};
 \node[xshift=2 cm,yshift=-1.5cm] (c2){  $4,5,6$};
 \node[xshift=-4cm,yshift=-3cm] (c3){ $1 $};
 \node[xshift=-1cm,yshift=-3cm] (c4){  $2,3$};
\node[xshift= 4cm,yshift=-3cm] (c6){ $6 $};
 \node[xshift= 1cm,yshift=-3cm] (c5){  $4,5$}; 
 \node[xshift= -2.0cm,yshift=-4.5cm] (c7){  $2$};
 \node[xshift= -0.5cm,yshift=-4.5cm] (c8){  $3$};
 
 \node[xshift= 0.5cm,yshift=-4.5cm] (c9){  $4$};
 \node[xshift= 2.0cm,yshift=-4.5cm] (c10){  $5$};
 \end{scope}

\draw[->, semithick] (c0) -- (c1);
\draw[->, semithick] (c0) -- (c2);
\draw[->, semithick] (c1) -- (c3);
\draw[->, semithick] (c1) -- (c4);
\draw[->, semithick] (c2) -- (c5);
\draw[->, semithick] (c2) -- (c6);

\draw[->, semithick] (c4) -- (c7);
\draw[->, semithick] (c4) -- (c8);
\draw[->, semithick] (c5) -- (c9);
\draw[->, semithick] (c5) -- (c10);

%\node[xshift=-4cm,yshift=-1.5cm] (c1){ Above average};
  
\node[xshift=8cm,yshift=-0.50cm] (c2){Query Agreement/Disagreement};
\node[xshift=8cm,yshift=-2cm] (c2){Query Extremity};
\node[xshift=8cm,yshift=-4cm] (c2){Query Weakness of Attitude};
\end{tikzpicture}
 \caption{A tree for six ordered categories, categories 1,2,3 represent levels of disagreement, categories 4,5,6 represent levels of agreement (compare Figure 3 in \citet{bockenholt2016measuring}).}
\label{tree1}
\end{center}
\end{figure}

\subsection{Binary IRTree Models}\label{sec:tree}

Tree-based models assume a nested structure with the building blocks given as binary models. They were considered, among others, by 
\citet{de2012irtrees},  \citet{bockenholt2013modeling}, \citet{khorramdel2014measuring},   \citet{bockenholt2016measuring} and \citet{bockenholt2017response}.
In the following we use the presentation of IRTree models given by \citet{bockenholt2016measuring}. IR-tree models are sequential process models, a response is constructed based on a series of mental questions. For illustration we consider an ordinal response with six categories  representing   ordinal outcomes ranging  from ``strongly disagree'' to ``strongly agree''. Figure \ref{tree1} shows the corresponding tree, which is equivalent to Figure 3 in \citet{bockenholt2016measuring}. The first query determines a respondent's agreement or disagreement. The second query determines the extremity of the (dis)agreement and the third query assesses whether the agreement is weak or not. For each query in the tree, which corresponds to a conditional binary decision one uses a binary model. For query $q$ 
the model is given by 
\begin{equation}\label{eq:query}
P(Y_{(q)pi}^{}=1)=F(\theta_p^{(q)}-\delta_i^{(q)}),
\end{equation}
and the (local) response variable $Y_{(q)pi}^{}$ is often referred to as a pseudo-item.

Pseudo-items are conditional dichotomizations, and can also be represented by split variables. 
For example, the query that determines the extremity within agreement categories, distinguishing between category 6 and categories $\{4,5\}$ corresponds to modelling the split variable $Y_{pi}^{(6)}|Y_{pi} \in \{4,5,6\}$, or alternatively $Y_{pi}^{(6)}|Y_{pi}^{(4)}=1$. Thus, tree models implicitly use the same
dichotomizations as traditional ordinal models.

However, there is one crucial difference between traditional models and IRTree models. While the former typically use one person parameter (and split-specific item parameters) the majority of IRTree models uses query-specific person parameters as given in (\ref{eq:query}).  This makes the models multi-dimensional in terms of person  parameters and person parameters are interpreted with reference to the specific query, that is, the conditional decision. In the tree given in Figure \ref{tree1} the basic propensity to agree or disagree is   modelled in the first query. The person parameters in the next queries refer to response styles, whether a person prefers extreme or middle categories. The parameterization seems not to efficiently use the information in the ordered categories 
since the propensity to agree or disagree is not present in later queries, though it might also determine the choice between category groups $\{1\}$ and $\{2,3\}$.

%Although the resulting model is rather flexible and easy to estimate a disadvantage is that for each query one has a new person parameter.   That makes the model multi-dimensional in terms of person  parameters and person parameters are interpreted with reference to the specific query, that is, the conditional decision. The model seems not to efficiently use the information in the ordered categories. In particular, the basic propensity to agree or disagree is only modelled in the first query. The person parameters in the next queries refer to response styles, whether a person prefers extreme or middle categories. However, the basic propensity to agree or disagree is not present in later queries though it might also determine the choice between category groups $\{1\}$ and $\{2,3\}$. In contrast,  a model  like the  sequential model uses the same person parameter in all binary decisions. Also in the partial credit model, which is not hierarchical, the same person parameter is present in the conditional binary decisions. 

Only recently more efficient binary trees have been proposed that use the same traits in more than one query \citep{TuDrax2019,meiser2019irt}.
In particular the approach of \citet{meiser2019irt} is very attractive. They do not simply use the same trait in different queries but use scaled versions, which may be seen as   factor loadings. Let, for example,  $\theta_p^{(1)}$ denote the trait in the first pseudo-item, which distinguishes between agreement and disagreement, then the pseudo item $Y_{(2)pi}^{}$, which distinguishes between  $\{1\}$ and $\{2,3\}$,  can be parameterized by 
\[
P(Y_{(2)pi}^{}=1)=F(\theta_p^{(2)} + \alpha\theta_p^{(1)} -\delta_i^{(2)}),
\] 
where the term $\theta_p^{(2)}$ represents the tendency to prefer categories $\{2,3\}$ (given the response is in categories $\{1,2,3\}$), and $\alpha\theta_p^{(1)}$ represents the scaled tendency to higher response categories. In a similar way scaled versions of traits from previous queries are used in other pseudo-items, for details see \citet{meiser2019irt}. The strength of such parameterizations is that the same person parameter is present on several levels of the tree, and parameters that are specific to pseudo-items get a distinct meaning, for example as a tendency to extreme or less extreme categories.  

\begin{figure}[h!]
\begin{center}
\begin{tikzpicture}[scale=.9, transform shape]
\begin{scope}[every node/.style={shape=ellipse,
    draw, align=center,top color=white, bottom color=blue!20}]
\node (c0){$1,2,3,4,5$};
\node[xshift=-2 cm,yshift=-1.5cm] (c1){ $1,2,4,5$};
 \node[xshift=2 cm,yshift=-1.5cm] (c2){  $3$};
  \node[xshift=-5cm,yshift=-3.5cm] (c3){ $1  $};
   \node[xshift=-3cm,yshift=-3.5cm] (c4){  $2$};
 \node[xshift= -1cm,yshift=-3.5cm] (c5){ $4$};
  \node[xshift= 1cm,yshift=-3.5cm] (c6){ $5$}; 
   %\node[xshift= -1.5cm,yshift=-4.5cm] (c7){  $5$};
  %\node[xshift= 2cm,yshift=-4.5cm] (c8){  $5$};
  %\node[xshift= 0.5cm,yshift=-4.5cm] (c9){  $4$};
 %\node[xshift= 2.0cm,yshift=-4.5cm] (c10){  $5$};
 \end{scope}
\draw[->, semithick] (c0) -- (c1);
\draw[->, semithick] (c0) -- (c2);
\draw[->, semithick] (c1) -- (c3);
\draw[->, semithick] (c1) -- (c4);
\draw[->, semithick] (c1) -- (c5);
\draw[->, semithick] (c1) -- (c6);
\node[xshift=8cm,yshift=-1.00cm] (c2){neutral or not};
\node[xshift=8cm,yshift=-2.5cm] (c2){within disagreement/agreement categories};
\end{tikzpicture}
 \caption{A tree for five ordered categories, categories 1,2  represent low response categories, categories 4,5  represent high response categories, 3 is the neutral middle category.}
\label{neutral}
\end{center}
\end{figure}
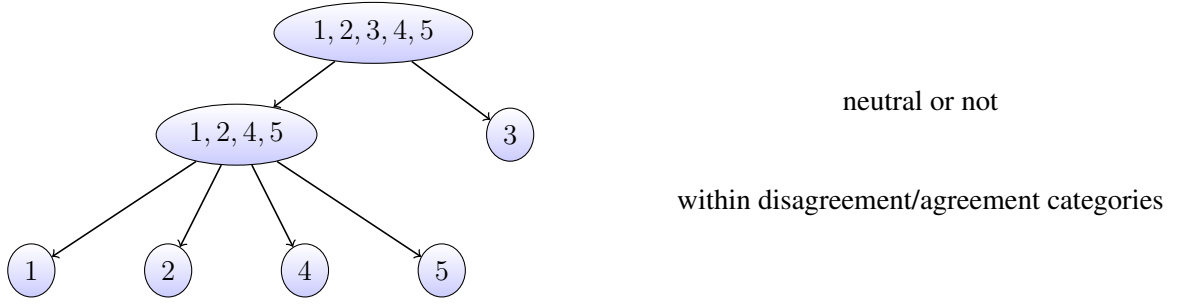

A major topic in binary trees is the modelling of response styles. However, IRTrees  provide  a wide class of flexible modelling tools that is not limited to response styles. For example, the first query may assess a person's tendency to select a midscale answer indicating neutrality in five or seven-grade Likert scales. Then the first split distinguishes between the middle category and other categories, the following splits model the response if the neutral middle category is avoided. The resulting tree has an asymmetric form, see  Figure \ref{neutral}. Models of this type are useful since the role of the neutral category is ambivalent.  \citet{kulas2008middle} investigated whether it is used to indicate a moderate standing on a trait/item, or rather is viewed by the respondent as a
'dumping ground' for unsure or non-applicable response. In the latter case the use of the middle category as part of the integer protocol
might yield strongly biased results. An initial binary split that distinguishes between the neutral category and other categories can avoid bias. Binary trees of this type have been considered by \citet{jeon2016generalized,bockenholt2016measuring}  and, more recently, by  \citet{plieninger2020developing,TuLikert2020}.

\subsection{Hierarchical Partitioning Using Ordinal Models}\label{sec:tree}

%Binary IRTrees can be seen as   special cases of hierarchically structured models. 
Binary splits are simple but yield rather large trees with many nodes. An alternative that exploits the ordering of categories and provides sparser parameterizations is to use ordinal models as building blocks.
%It might be attractive to use simpler structured hierarchical models that use the order in response categories within the tree.  
 
\begin{figure}
\begin{center}
\begin{tikzpicture}[scale=.9, transform shape]

%\begin{scope}[every node/.style={shape=rectangle, rounded corners,    draw, align=center,top color=white, bottom color=blue!20}]
\begin{scope}[every node/.style={shape=ellipse,     draw, align=center,top color=white, bottom color=blue!20}]

\node (c0){$1,2,3,4,5,6$};
\node[xshift=-2 cm,yshift=-1.5cm] (c1){ $1,2,3$};
 \node[xshift=2 cm,yshift=-1.5cm] (c2){  $4,5,6$};
 \node[xshift=-4cm,yshift=-3cm] (c3){ $1 $};
 %\node[xshift=-1cm,yshift=-3cm] (c4){  $2,3$};
\node[xshift= 4cm,yshift=-3cm] (c6){ $6 $};
 %\node[xshift= 1cm,yshift=-3cm] (c5){  $4,5$}; 
 \node[xshift= -2.0cm,yshift=-3cm] (c7){  $2$};
 \node[xshift= -0.5cm,yshift=-3cm] (c8){  $3$};
 
 \node[xshift= 0.5cm,yshift=-3cm] (c9){  $4$};
 \node[xshift= 2.0cm,yshift=-3cm] (c10){  $5$};
 \end{scope}

\draw[->, semithick] (c0) -- (c1);
\draw[->, semithick] (c0) -- (c2);
\draw[->, semithick] (c1) -- (c3);
 \draw[->, semithick] (c1) -- (c7);
 \draw[->, semithick] (c1) -- (c8);
\draw[->, semithick] (c2) -- (c6);

%\draw[->, semithick] (c4) -- (c7);
 \draw[->, semithick] (c2) -- (c9);
 \draw[->, semithick] (c2) -- (c10);
%\draw[->, semithick] (c5) -- (c10);

%\node[xshift=-4cm,yshift=-1.5cm] (c1){ Above average};
  
\node[xshift=8cm,yshift=-0.50cm] (c2){Query Agreement/Disagreement};
\node[xshift=8cm,yshift=-2cm] (c2){Query Extremity};
%\node[xshift=8cm,yshift=-4cm] (c2){Query Weakness of Attitude};
\end{tikzpicture}
 \caption{A tree for six ordered categories with three levels .}
\label{hier22}
\end{center}
\end{figure}
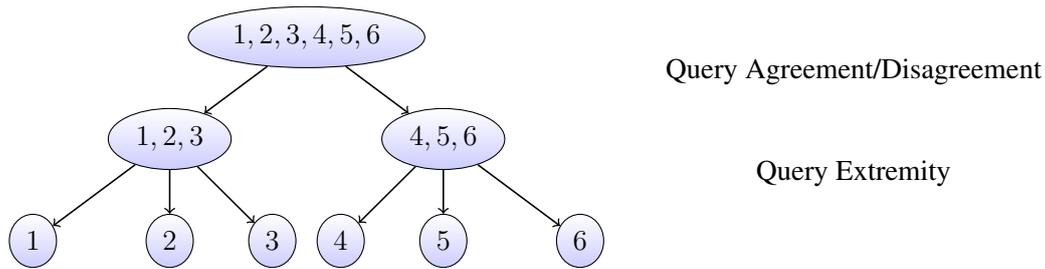

Let us again consider an example with  six ordered categories. Instead of using the binary splits tree given in Figure \ref{tree1} one can work with  the tree given in Figure \ref{hier22}. It has a simpler structure with  only two levels in addition to the 0-level, which contains all categories.
One can model the propensity to agree or disagree by 
\begin{equation}\label{eq:tree}
P(Y_{pi} \ge 4)=F(\theta_p-\delta_i^{(1)}),
\end{equation}
where $\delta_i^{(1)}$ is the  level 1 item parameter. The conditional propensity to choose from one of the categories in level 2 can be specified by any simple ordinal model, for example, by  conditional graded response models,
\begin{align}\label{eq:tree2}
&P(Y_{pi}\ge r|Y_{pi} \le 3)=F(\alpha\theta_p-\delta_{ir}^{(2)}), r= 2,3, \\ &P(Y_{pi} \ge r|Y_{pi} \ge 4)=F(\alpha\theta_p-\delta_{ir}^{(2)}), r= 5,6,
\end{align}
where $\alpha$ scales the person parameter at the second level. The model has just one parameter more than the simple graded response model, however, order restrictions are weaker. One just has $\delta_{i2}^{(2)} \le \delta_{i3}^{(2)}$
and $\delta_{i5}^{(2)} \le \delta_{i6}^{(2)}$ whereas in the simple cumulative model five thresholds have to be ordered.
\citet{thissen2013two} considered a two-decision model of this type, which uses a modified graded response model in the second level.

Within the model it is straightforward to include response styles by adding just one person parameter. In the extended model
%\begin{align}\label{eq:tree2}
%&P(Y_{pi}\ge r|Y_{pi} \le 3)=F(\alpha\theta_p+\gamma_p-\delta_{ir}^{(2)}), r= 2,3, \\ &P(Y_{pi} \ge r|Y_{pi} \ge 4)=F(\alpha\theta_p+\gamma_p-\delta_{ir}^{(2)}), r= 5,6,
%\end{align}
$P(Y_{pi}\ge r|Y_{pi} \le 3)=F(\alpha\theta_p+\gamma_p-\delta_{ir}^{(2)})$, $r= 2,3$, $P(Y_{pi} \ge r|Y_{pi} \ge 4)=F(\alpha\theta_p-\gamma_p-\delta_{ir}^{(2)})$,  $r= 5,6$, the parameter $\gamma_p$ is a response style parameter that contains  the tendency to middle categories. In a binary
tree as given in Figure \ref{tree1} several additional parameters are needed to account for response styles whereas in the simpler structured tree in Figure  
\ref{hier22} only one additional parameter is needed. For the estimation one can use similar methods as in binary trees,  exploiting that likelihood contributions can be written as products of conditional probabilities  \citep{TuDrax2019}. 

The class of hierarchical partitioning models is characterized by containing ordinal models for more than two categories as constituents. Instead of using just binary splits the order in responses is exploited efficiently by using ordinal models as building blocks. When modeling the response within agreement and disagreement categories any simple ordinal model can be used. Since the ordinal models can be represented by split variables the same holds for the model built from these blocks. They are in particular helpful to obtain   sparse parameterizations.

%In general, after the first step hierarchical approaches model the response as a sequence of conditional responses, they are \textit{process models}.  

%Hierarchical partitioning models are a wide class of models, which can be tailored to specific settings.  

%Although they are not directly constructed from binary models they can be seen as containing dichotomizations since the ordinal building blocks can be represented by split variables.  
%modeling in the first step the response in groups of homogenous response categories and then modeling the response within groups. If the number of agreement and disagreement categories is even it is natural to split in the first step into categories $\{1,\dots, m/2\}$ and $\{m/2+1,\dots, m\}$ by utilizing a binary model. If the number of response categories is odd with  a neutral category in the middle one can use a three categories model in the first step. 

%%%%%%%%%%%%%%%%%%%%%%%%%%

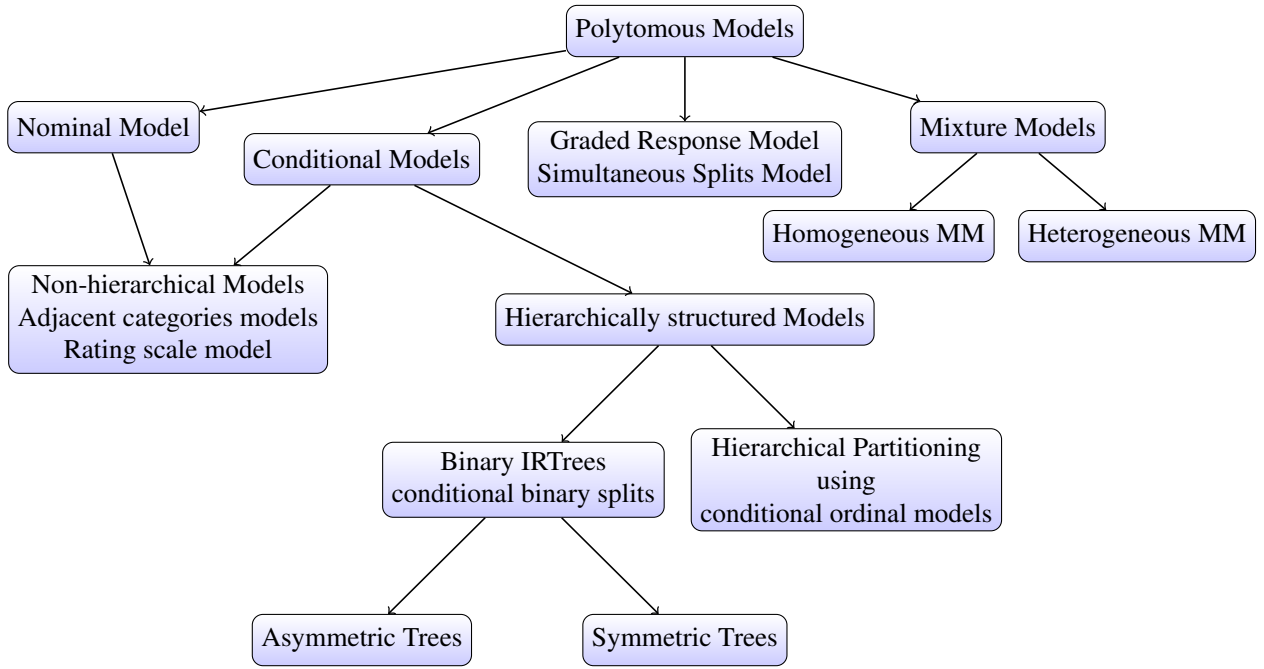
\begin{figure}
\begin{center}
\begin{tikzpicture}[scale=.85, transform shape]

\begin{scope}[every node/.style={shape=rectangle, rounded corners,
    draw, align=center,top color=white, bottom color=blue!20,minimum height=0.8cm}]
\node[xshift=-1.0 cm,yshift=0cm] (c0){Conditional Models};
\node[xshift=-4 cm,yshift=-2.5cm] (c1){ Non-hierarchical Models\\Adjacent categories models\\Rating scale model};
 \node[xshift=4 cm,yshift=-2.5cm] (c2){ Hierarchically structured Models};
 %\node[xshift=-1.5cm,yshift=-5cm] (c3){ Sequential Model\\compares category and groups};
  \node[xshift=1.5cm,yshift=-5cm] (c4){ Binary IRTrees\\conditional binary splits};
 \node[xshift=6.5cm,yshift=-5cm] (c6){ Hierarchical Partitioning\\using\\conditional ordinal models};
\node[xshift=4cm,yshift=0cm] (c5){Graded Response Model\\Simultaneous Splits Model};
\node[xshift=9cm,yshift=0.5cm] (c7){Mixture Models};
\node[xshift=7.0cm,yshift=-1.2cm] (c8){Homogeneous MM};  
\node[xshift=11cm,yshift=-1.2cm] (c9){Heterogeneous MM};
 \node[xshift=4cm,yshift=2cm] (c00){Polytomous Models}; 
 \node[xshift=-1.0cm,yshift=-7.5cm] (c10){ Asymmetric Trees};
 \node[xshift=4.0cm,yshift=-7.5cm] (c11){ Symmetric Trees};
 \node[xshift=-5.0cm,yshift=0.5cm] (c01){ Nominal Model};
   
  \end{scope}
\draw[->, semithick] (c00) -- (c01);
\draw[->, semithick] (c01) -- (c1);
\draw[->, semithick] (c0) -- (c1);
\draw[->, semithick] (c0) -- (c2);
%\draw[->, semithick] (c2) -- (c3);
 \draw[->, semithick] (c2) -- (c4);
 \draw[->, semithick] (c2) -- (c6);
\draw[->, semithick] (c00) -- (c0);
 \draw[->, semithick] (c00) -- (c5); 
%\draw[->, semithick] (c4) -- (c3);
\draw[->, semithick] (c00) -- (c7);
\draw[->, semithick] (c7) -- (c8);
\draw[->, semithick] (c7) -- (c9);
\draw[->, semithick] (c4) -- (c10);
\draw[->, semithick] (c4) -- (c11);
%\node[xshift=8cm,yshift=-2cm] (c2){Query Extremity};
%\node[xshift=8cm,yshift=-4cm] (c2){Query Weakness of Attitude};
\end{tikzpicture}
 \caption{Hierarchy of polytomous  models.}
\label{hier2}
\end{center}
\end{figure}
%%%%%%%%%%%%%%%%%%%%%%%%%%%%%%%%%%%%%%%%%%%%%%%%%%%%5

\section{A Taxonomy of Polytomous Item Response Models Including Tree Structured Models}\label{sec:tax}

The taxonomy of ordinal models given in Figure \ref{struc1} covers only  the basic models.  An extended taxonomy of polytomous IRT models that also contains the general class of hierarchically structured models is given in Figure \ref{hier2}. It also includes the  nominal  model and mixture models to be  considered later %(Section \ref{sec:NomPar}). 
Here we focus on the structure that is obtained 
by they way how  ordinal models can be constructed from  binary  models as building blocks.     
At the outset it is distinguished between conditional models and simultaneous splits, that is, graded response  models. The former use binary models in a conditional way  by assuming that the choice between categories has already been narrowed down to a reduced set of categories. In contrast, the latter assume no conditioning but assume that the splits between categories are simultaneously determined by the same person parameter. 

There are two groups of conditional models. In the first group  pairs of categories are compared by utilizing a binary response model to obtain, for example, the partial credit model and its simplified or extended versions. The second group is formed by hierarchical models. The crucial difference between non-hierarchical and hierarchical models is that in the former the   conditions under which   binary models are assumed to hold are overlapping. For example, in the partial credit model one binary  sub model conditions on the the categories $\{0,1\}$, another sub model conditions on $\{1,2\}$. Both conditions contain the category $1$. This overlapping  prevents a representation as a hierarchical model.

Hierarchically structured models can be divided into two types of models, binary IRTrees and hierarchical partitioning approaches. The former use only binary models to describe the conditional response in subsets of categories while the latter use traditional models with more than two categories as building blocks. 

One can further distinguish between symmetric and asymmetric tree models. Symmetric models are in particular useful for Likert items to account for the symmetry in answer categories. \textit{Symmetric tree models} can be defined by considering subsets of categories $S_1, S_2 \subset \{1,\dots, a\}$, where $a=m/2$ if $m$ is even, and $a=(m-1)/2$ if $m$  
is odd.  An IRTree model is symmetric if for any (conditional) split between $S_1$ and $S_2$ there is a split between $S_3=\{r|m-r+1, r\in S_2\}$ and $S_4=\{r|m-r+1, r\in S_1\}$. An example with an even number of categories is the splitting structure shown in Figure \ref{tree1}. If the number of categories is odd, and Likert items are considered, the first split  typically distinguishes between the neutral category and the other categories. Although the visual appearance of the corresponding tree shows some asymmetry, see Figure \ref{neutral},  the corresponding model is symmetric, and treats categories in a proper way.
Therefore, it is essential to distinguish between the symmetry of a tree and the symmetry of the model. While the former refers to the tree structure the latter refers to the corresponding model.

A classical example of a tree model that is not symmetric is the 
sequential model. It is, in particular,  not invariant under the reverse permutation of categories; if the order of categories is reversed the corresponding sequential model differs from the sequential model for the original categories. In contrast, most   symmetric models in common use are invariant under the reverse permutation, namely symmetric models,  in which the response function $F(.)$ is a symmetric distribution function. The distinction between asymmetric and symmetric models can be made for all hierarchically structured models. It is included in the taxonomy only for binary IRTrees, which have been investigated in the literature more intensively than other hierarchical models.

In general, the graded response model and the adjacent categories models can be used for any form of graded responses, in achievement tests as well as in the investigation of attitudes. Hierarchically structured models are somewhat different, they are process models   tailored to model a specific process. The sequential model assumes that levels of performance are reached successively, and therefore is most useful in items that are constructed   with categories that represent successive solutions levels. Binary IRTrees and hierarchical partitioning approaches  assume a specific conditional structure that aims at modeling the way how respondents generate a response. In hierarchically structured models, as in all conditional models, item parameters have to be interpreted locally since they refer to conditional decisions.

%asymmetric and symmetric models. The sequential model assumes a sequential process and therefore conditions on the level that has been reached.   IRTrees are general hierarchical models that allow almost arbitrary binary splits. They seem useful in particular for Likert items in which the categories are divided into agreement and disagreement categories. Of course, if one does not restrict the construction to Likert type items the sequential model can be considered a special case of   general IRTree models, which is visualized by the arrow that connects the two types of models. Nevertheless it is  a very specific hierarchically structured model, in particular it is asymmetric in contrast to most IRTree models used to model Likert items. Hierarchical partitioning models are general models that contain   ordinal models in the split levels. 
 
%In specific cases, for example, if one has only four categories in a disagreement - agreement item, of course all the splits are binary and one obtains an IRTree model. 

%The structure proposed here has the purpose of characterizing modeling approaches and describing relationships.  The tree structure in Figure \ref{hier2}  shows the branching within the class of polytomous models. It is in particular seen that a graded response model cannot be represented as a partial credit model, and the partial credit model is not a hierarchically structured model. 

\section{Nominal Models, Ordinal Models and the Role of Parameterizations}\label{sec:NomPar}

All the polytomous IRT models considered so far can be considered ordinal models in the sense that they exploit the ordering of categories.
A model that is different in this aspect is the so-called nominal model. It can be seen as a model that aims at detecting the order rather than using it, but also as a sort of background model from which specific ordinal  models can be derived. In a taxonomy of polytomous IRT models, which essentially is a taxonomy of ordinal models, it should  be included and its role be investigated.

Another major   topic in the following is the role of the parameterization of a model, which also plays a role in the transformation of the nominal model into an ordinal model.  Variations in parameterizations yield   more or less complex models of specific model types  in the hierarchy given in Figure \ref{hier2}, various parameterizations can be used on every level of the hierarchy (Section \ref{sec:param}).    
Another aspect of the parameterization within the taxonomy considered here concerns the link between parameterization and the exploitation of the ordering of categories. It is argued that  the split variables and their specific parameterization make models   ordinal models. 
%This link makes models ordinal models.    

\subsection{Nominal Models}\label{sec:nominal}
The taxonomy  uses that ordinal models can be seen as composed from simpler, in particular binary models. This is most obvious in IRTrees but 
holds also for basic models as the graded response model. What makes the models ordinal ones is that the binary models are assumed to hold only for \textit{specific} subsets of categories. For example, the graded response model assumes binary models to hold for subsets $\{0,\dots,r\}$ and  $\{r+1,\dots,k\}$. None of the ordinal models are built from  binary models that distinguish between  subsets such as $\{3,7\}$ and $\{5\}$. All subsets that are used  reflect the ordering of the categories $\{0,\dots,k\}$. More concrete, binary models are assumed to hold, possibly conditionally, for subsets $S_1, S_2$, with $c_1 < c_2$ for $c_1 \in S_1, c_2 \in S_2$, and the binary models distinguish between $S_1$ an $S_2$ in a way such that an increase in $\theta_p$ increases the probability of a response in $S_2$. It can be seen as an 'ordered  subsets' characterization of ordinal models, which is linked to split variables as considered later.

%This characterization is linked to the split variables defined in  , which can be used to define when a model uses the ordering.

This aspect is emphasized since ordinal models are sometimes derived  from models that do not use the order of categories. The most widely used model to this end is  Bock's nominal model \citep{bock1972estimating} %(in Thissen/Cai, Handbook, wrong in the book of Mair) is 
\begin{equation}\label{eq:bock}
P(Y_{pi}=r)= \frac{\exp(\alpha_{ir} \theta_p-\beta_{ir} )}{\sum_{s=0}^{k}\exp(\alpha_{is} \theta_p-\beta_{is} )}, \quad r=1,\dots,k,
\end{equation} 
in which additional constraints are needed to ensure identifiability of parameters, see \citet{bock1972estimating}, \citet{ThissCai2016}. In the basic form the model uses only the nominal scale of the response, however, it can be transformed to use the order information. A first step is to set $\alpha_{ir}=\phi_r$, where $\phi_r$ are considered scoring functions for the 
categories, yielding Andersen's version of the model \citep{Andersen:77}. If,   in addition, it is assumed that the scores are ordered, that is, $\phi_1 \le \dots \le \phi_k$, one obtains a model that actually uses the ordering of categories. If one assumes equi-distant scores, $\phi_r=r$ one obtains the partial credit model, which has been noted  among others by \citet{thissen1986taxonomy}, where much more general transformations were considered. 
Also a  general-purpose multidimensional  model was considered by \citet{thissen2010nominal}.

%The point that is emphasized here is that the nominal model can be considered a background model from which the partial credit model can be derived as a special case, obtained by a very specific scaling of categories. This is  theoretically  interesting but it should be noted that the nominal model, although being a model for nominal categories, is somewhat peculiar in its basic form. It is a model that contains a \textit{uni-dimensional} person parameter, however, uni-dimensionality is somehow contradictory to the purpose of modelling nominal response categories. If categories are measured on a nominal scale it is much more sensible to allow for multi-dimensional person parameters, which were considered by \citet{thissen2010nominal}. Also multinomial models derived as random utility models assume multi-dimensional utilities, see \citet{Yellott:77,McFadden:73}, as do nested logit models \citet{McFadden:81}. That means, the nominal model in its basic form seems not a sensible model, it only becomes a useful model by using specific scales or constraints,  or by extending it to multi-dimensional traits.

The nominal model can be seen as a useful background model from which various ordinal models can be derived as special cases. It is also interesting from a conceptual point of view since it  has also been used in a different way, namely to check the order of categories if that is not clear. This use is linked to different conceptualizations of ordinal and nominal models. An ordinal model, in the sense used here,  exploits the order of categories while a nominal model is a model that is invariant against permutations of response categories. In its general form the  model (\ref{eq:bock})  is  a nominal model but not an ordinal one since it is stable under permutations.

On the other hand it uses a uni-dimensional trait, which implicitly assumes an order of the latent trait and therefore on the responses. This makes it a model that can be used to investigate the order of categories. Fitting the unconstrained model might yield information on the order, and it can be used to fit responses constructed for testlets, see, for example, \citet{ThissCai2016}.  It can also be used to provide scores using information from all responses even when the response categories are not clearly ordered. When used in this way it does not exploit the order of categories but aims at investigating the order empirically. This is a different concept of dealing with ordinality, namely using the model 'to examine the expected, or empirical, ordering of response categories' \citep{ThissCai2016}.

It becomes an ordinal model in the sense used here if restrictions are imposed. Moreover, it generates a whole family of models that is strongly linked to the partial credit model. It should be noted that it does not generate the general adjacent categories model, but only models that use the logistic link, which, however, are the most widely used ones.  This is visualized in the tree structure given in  Figure \ref{hier2}.
 Adjacent categories models are sub models of nominal models (if the logistic link is used) but can also be considered as specific conditional models (for any response function $F(.)$).
%The nominal model could be included to obtain a (slightly enlarged) taxonomy of item response models. However, it is linked to the adjacent categories model in the form of the partial credit model only. It is not even a model that comprises   adjacent categories models that do not use the logistic distribution function. Thus, not much is gained by including it in the structure shown in Figure \ref{hier2}. The essential variety of models  is due to the use of the ordering of categories, which is reflected in the given taxonomy. 

%The uni-dimensionality makes it a very special nominal model. Classical models for nominal responses as the multinomial logistic choice model assume multi-dimensional utilities, one for each alternative. The observed response is considered to be the one following from the principle of maximum random utility, which means that that response is observed  that corresponds to the maximal random utility, see \citet{Yellott:77,McFadden:73}, the same holds for the more general nested logit models \citet{McFadden:81}.

\subsection{Models and Parameterizations}\label{sec:param}

In the proposed taxonomy  the parameterization is considered secondary. That does not mean that parameterization is not important. It is    important, and much of the more recent latent trait literature is devoted to account for specific features like response styles or  differential item functioning, which can be investigated by using specific parameterizations. However, parameterizations do not alter the structure given in Figure \ref{hier2}. They may be seen as special cases within this framework.

Given the conditional or unconditional structure of a model  quite differing parameterizations can be used. Instead of the simple difference between person parameters $\theta_p-\delta_{ir}$ one can include slope parameters yielding $\alpha_{i}(\theta_p-\delta_{ir})$. Response style effects can be modeled by adding additional person parameters yielding models with a multi-dimensional person parameter, see \citet{johnson2003use} for cumulative type models with extreme response styles, and \citet{wetzel2015multidimensional},  \citet{plieninger2016mountain},   \citet{jin2014generalized}, \citet{TuSchBe2018} for  partial credit models that account for  response styles. 
  
The taxonomy given in Figure \ref{hier2} also includes models that make much weaker assumptions on the response functions.  Introduced by   \citet{mokken1971theory} nonparametric IRT models have been extended to a wide class of models nonparametric IRT models, see, for example, \citet{sijtsma2002introduction}. Assumptions are much weaker, only local independence, uni-dimensionality and some form of monotonicity are needed. Ordinal nonparametric models can be derived  by using more general functions in the binary models that are the building blocks of the models in the taxonomy. Instead of using the parametric form $F(\theta_p-\delta_{i})$  one uses a uni-dimensional monotonic function. For example, the cumulative version is obtained by assuming   
$P(Y_{pi} \ge r)= M_{ir}(\theta_p)$,
where $M_{ir}$ is a strictly increasing function that can depend on the item $i$ and the response category $r$. Corresponding adjacent categories and sequential models are obtained by using on the left hand side $P(Y_{pi} = r | Y_{pi} \in \{r-1,r\})$ or $P(Y_{pi} = r | Y_{pi} \ge r-1)$, respectively.
Models of this form have  been considered by \citet{hemker1997stochastic, hemker2001measurement}. 

Parameterizations yield specific hierarchies if the conditioning (the type of model) and the response function are fixed. An example is the hierarchy given by \citet{hemker2001measurement} in their Figure 2 for sequential models, which starts with   the very restrictive  sequential rating scale model, in which the location parameter is split into an item location parameter and a step parameter. The most general models in this hierarchy are Samejima's acceleration model 
\citet{samejima1995acceleration} and the nonparametric sequential model. Similar hierarchies may be built for all of the models identified in Figure \ref{hier2} but they are hierarchies generated by parameterization within the taxonomy. The taxonomy itself, which shows the relationship of models characterized by their   conditioning, is unchanged.

Taxonomies that focus on parameterizations have been given by  \citet{hemker1997stochastic}, \citet{hemker2001measurement} and \citet{sijtsma2000taxonomy}.  They study carefully which parameterizations are special cases of other ones and display the structure in  Venn diagrams.  They also investigate  so-called measurement properties of models as the monotone likelihood ratio, stochastic ordering properties and invariant item ordering, and show which models have these properties.

\subsection{Characterization of Ordinal Models}

In the following the fundamental structure of ordinal models is investigated. It is argued that binary models for split variables are the essential constituents of models that are able to exploit the ordering of categories. Although typically there is some intuition why models are appropriate for ordered responses, for example the ordered thresholds on the latent scale in cumulative models, and the process from which the sequential model is derived, these motivations do not yield a general conceptualization of ordinal models. Nevertheless, in a taxonomy of polytomous models it seems warranted that one  distinguishes between ordinal and nominal models.

%Let us come back to the question how    an ordinal model can be characterized in general. In the following it is argued that the parameterization is not what makes a model an ordinal one. It is linked to  the split variables, which can also be seen as being behind the 'ordered  subsets' characterization of ordinal models considered when discussing nominal models.

%Let us consider the general graded response model $P(Y_{pi} \ge r)=F(\theta_p^{r}-\delta_{ir})$, where the person parameter is category specific. That means one has a multidimensional traits model 

%Let us first consider, which specific binary models are contained in polytomous models. 
All the models considered here contain binary submodels of the form $P(Y_{p+} \in S_1| Y_{p+} \in S_2)= g(\theta_p, \{\delta_{is}\})$, where $S_1 \subset S_2$ and $\{\delta_{is}\}$ is a set of item parameters. For example, the tree in Figure \ref{tree1} contains a binary model that distinguishes between $\{1\}$ and $\{2,3\}$ given $\{1,2,3\}$. The corresponding  model for $P(Y_{pi} \in \{2,3\}| Y_{pi} \in \{1,2,3\})$ can also be described by split variables, it is equivalent to modelling $P(Y_{pi}^{(2)} =1| Y_{pi}^{(4)} =0)$. Also in partitioning models that have ordinal components binary models for split variables are contained. If a cumulative model is used to model the response given $\{1,2,3\}$ in the structure given in Figure \ref{hier22} the model contains a binary model   $P(Y_{pi} \in \{2,3\}| Y_{pi} \in \{1,2,3\})$, which is equivalent to $P(Y_{pi}^{(2)} =1| Y_{pi}^{(4)} =0)$. 
%In general, a polytomous response model can be considered  an \textit{ordinal model}  if there exists a finite number of submodels  

However, binary models for split variables are not only submodels of ordinal model but are also  the building blocks of the models. Therefore, a general class of models   can be  defined by postulating that  there exists a finite number of submodels for split variables
\begin{equation}\label{eq:subm}
P(Y_{pi}^{(l)}=1|  Y_{pi}^{(s)}=1, Y_{pi}^{(r)}=0)= g_l(\theta_p^{(l)}| \{\delta_{is}\}), \quad   s< r,
\end{equation} 
such that 
\begin{itemize}
\item[(i)]the function $g_l(.|\{\delta_{is}\})$ is nondecreasing for at least one $l$ with $s \le l \le r$, 
\item[(ii)] the response probabilities are uniquely determined by these submodels.
\end{itemize}
The set of submodels are called the \textit{model generating binary models}, and the class of generated models  \textit{split variables generated ordinal models}. 

The second condition just ensures that the polytomous model can be constructed from the set of  binary models. The  crucial ingredients in the definition  are the form of the conditioning in (\ref{eq:subm}) and that the function is  nondecreasing. 
The condition $Y_{pi}^{(s)}=1, Y_{pi}^{(r)}=0$   means that conditioning  refers to a \textit{sequence of categories}  since it is equivalent to  
$Y_{pi} \in \{s, s+1,\dots,r-1\}$. It can also be empty, which is the case if $s=0, r=k+1$, and one defines $Y_{pi}^{(k+1)}=0$.  
The postulate that the function is nondecreasing ensures that the order of categories is used in a consistent way. It implies that, whatever the conditioning, an increase in $\theta_p^{(l)}$ is in favor of higher categories, the probability of lower categories can not increase. It excludes, in particular, that one constructs a binary tree, in which, for example, the response in the lower category $1$ given $\{1,2\}$  and the response in the higher category 6 given $\{5,6\}$ are modeled by  binary Rasch models. This construction would violate the order of categories, and is avoided  by using split variables to define the condition. 

%Essential ingredients are that functions are nondecreasing and that split variables, which represent the order, are constituent elements.  
%Therefore the ordinal models uses that ordinal models can be constructed from binary submodels, which in turn are contained in the model. 
With the exception of the nominal model, all models considered here are split variables generated ordinal models. In the traditional models, that is, the cumulative, sequential, and adjacent categories model, it is typically assumed that the trait does not depend on the the split, that means, one has in all models generating binary models with $\theta_p^{(l)}=\theta_p$ for all $l$. The model generating binary models are the ones given in Table \ref{tab:over}. 
In IRTrees the parameters $\theta_p^{((l))}$ are  not necessarily the same, they can vary over the binary models.

It is instructive to investigate why the nominal model is not among the class of models specified by (\ref{eq:subm}). In particular, it clarifies the conditions in the definition of this class of models. 
The nominal model (\ref{eq:bock})  is certainly a nominal but not an ordinal model since it is invariant under permutations. Nevertheless, it can be constructed from binary submodels $\log (P(Y_{pi}=r)/P(Y_{pi}=1)) =\gamma_{ir} \theta_p-\xi_{ir}$  with $\gamma_{ir} \ge 0$. These submodels yield  $\log (P(Y_{pi}=r)/P(Y_{pi}=r-1)) =(\gamma_{ir}-\gamma_{i,r-1}) \theta_p+\xi_{i,r-1}-\xi_{i,r}$. This model is equivalent to the nominal model, which is seen by using the  reparameterization  $\alpha_{ir}=\gamma_{ir}-\gamma_{i,r-1}$, $\beta_{ir}=\xi_{i,r}-\xi_{i,r-1}$.   Thus, a  nominal model is constructed from binary models that are  nondecreasing in person parameters because $\gamma_{ir} \ge 0$ is assumed. The crucial point is that the model generating submodels $\log (P(Y_{pi}=r)/(P(Y_{pi}=1) =\gamma_{ir} \theta_p-\xi_{ir}$ are based on the conditioning $Y_{pi} \in \{1, r\}$, which is not of the type used in the definition of the class of 
split variables generated ordinal models. This clarifies that one has to postulate for the (conditional) model generating binary models that the condition is of the form $Y_{pi}^{(s)}=1, Y_{pi}^{(r)}=0$  or, equivalently, $Y_{pi} \in \{s, s+1,\dots,r-1\}$. Without that condition it would not be ensured that a model  exploits the ordering of categories.

The class of split variables generated ordinal models  comprises all the traditional ordinal models as well as the hierarchical models considered previously. It   does not depend on specific parameterizations, it is just assumed that response functions are nondecreasing, and that the condition in the generating model has a specific form determined by split variables.  Although it can not be totally excluded that there might be alternative ways to find models that use the order in categories, the considered class of models seems rather exhaustive.

An additional advantage of characterizing  ordinal models by the  binary submodels that are contained is that it is rather flexible and avoids questionable criteria. For example, \citet{adams2012rasch} consider categories in an item response model as ordered if the expectation $\E(Y_{p}|\theta_p)$ is an increasing function of $\theta_p$. That means, if a person has a higher value of $\theta_p$ than another person, then the person with the higher value will, on average, score more.  The problem with the definition is that the expectation is a sensible measure only if the response $Y_{p}$ is measured on a metric scale level, however, one wants to characterize the use of the ordinal scale level. The expectation is   not helpful for this purpose because it uses a scale level that is not assumed to be available. Alternative ways of characterizing the use of order rely on specific functions. As \citet{adams2012rasch} suggest a model uses the order if for any ordered pair $s<r$ the function $m_{sr}(\theta_p)={P(Y_{pi} = r |  \theta_p)}/{P(Y_{pi} = s |  \theta_p)}$  is an increasing function  of $\theta_p$. It can   be seen as a  {stochastic ordering} property since it implies that for $\theta_{p_1}< \theta_{p_2}$ one has $m_{sr}(\theta_{p_1})<m_{sr}(\theta_{p_2})$.  However,  one might use quite different functions, for example, that $P(Y_{p} \ge r |  \theta_p)$  is an increasing function  of  $\theta_p$, yielding quite different conceptualizations of  ordinal models that are not compatible. The strength of the conceptualization based on split variables is that no specific functions, which are somewhat arbitrary, are needed. 

% \citep{TuWhat2019}. Functions of this form seem not to provide general criteria for characterizing ordinal models since there is a variety of ways to define stochastic ordering. Interestingly the alternative ways to define order by stochastic ordering considered in \citet{TuWhat2019} are in parts parallel to the alternative forms of stochastic ordering based on sum scores as considered by \citet{hemker2001measurement, van2005stochastic}, which have been discussed above.

%\subsubsection*{Collapsibility}

\section{Mixture Models}\label{sec:mix}
An alternative class of models that has been included in Figure \ref{hier2} are mixture models. They follow a quite different reasoning to account for heterogeneity in responses and response styles and therefore are included as a separate class of models. General finite mixture models for latent traits have the form
\[
P((Y_{p1},\dots,Y_{pI}))= \sum_{m=1}^{M} \pi_m  P_m((Y_{p1},\dots,Y_{pI})|\theta_p, \{\deltab_{ir}^{(m)}\}).
\]
That means the population is subdivided into $M$ latent classes, where 
$P_m(.)$ denotes  the model in the latent class $m$ with parameters $\theta_p, \{\deltab_{ir}^{(m)}\}$, and $\pi_1,\dots,\pi_M$ denotes the mixture probabilities of the latent classes ($\sum_m \pi_m=1$).

Mixture item response   models, originally developed for Rasch models by \citet{rost1991logistic}, are strong tools to investigate unidimensionality, the presence of response styles,   
and differential item functioning without assuming that the relevant grouping variable that induces differential item functioning to be known.
Extensions to ordinal responses have been considered by \citet{rost1997applying}, \citet{eid2000detecting}, \citet{gollwitzer2005response}, \citet{maij2008fitting}, \citet{Moors2010},\citet{VanRos2010}, \citet{von2004partially}, for an overview see also \citet{von2007mixture}.

It seems sensible to distinguish between two approaches to specifying mixture models, the homogeneous modelling strategy and the heterogeneous strategy. 
In \textit{homogeneous finite mixture models} the  same functional form is used in all the mixture components, for example, a partial credit model  \citep{eid2000detecting,gollwitzer2005response}.  It is assumed that respondents are from different latent classes but only model parameters, not the structure of the model vary across classes. The approach is not without problems. Typically  
the number of classes is unknown and has to be chosen driven by data. However, one  gets quite different model parameters when fitting, for example,   three or four classes, since all the parameters   change when considering one more class. Even if a number of classes has been chosen it is sometimes still difficult to interpret the difference between classes and  explain what exact features are represented by  classes, they might indicate a response style or some other dimension that is involved when responding to items. Homogeneous models do not explicitly model which trait is to be detected and are primarily  exploratory tools. 

\textit{Heterogeneous finite mixture models} are sharper tools, they allow to use different models in the components specifying  explicitly which specific trait is to be detected. Moreover, typically the number of components is fixed. Early mixture models of this type are HYBRID models as proposed  by  
\citet{von1996mixtures,von2007mixture}. Although some HYBRID models can be represented as mixture models that have the same functional form in the components 
but with constraints in some of the components \citep{von2007mixture}, the constraints specify which traits are modelled. Further models with constraints have been proposed by \citet{de2011explanatory}, \citet{shu2013using}. 

A specific mixture  with fundamentally different components as been considered more recently by  \citet{tijmstra2018generalized}. They proposed a two-class mixture of a generalized partial credit model and an IRTree model, carefully designed to distinguish between respondents who consider the middle category of a five-category Likert item as representing one category in a sequence of ordered categories and respondents who use the middle category as a non-response option. While the former follow a partial credit model the response of the latter is described by a specific IRtree model that separates the middle category.

All mixture models, in which at least one of the components is an ordinal model account for the order of categories. Therefore, they are included in the taxonomy,
but they are a separate class of models with specific purposes. In particular they can be used to model response styles in a quite different way than IRtrees and extensions of classical models as the partial credit model with response style. In the following possible approaches are considered briefly.

To avoid the pitfalls of homogeneous mixture models it might be sensible to use   structured mixtures, in which the type of response style is explicitly specified. With response vector $\Yb_p^T=(Y_{p1},\dots,Y_{pI})$ a simple two-components model has the form 
\[
P(\Yb_p)=  \pi_M  {P_M(\Yb_p|\theta_p, \{\deltab_{ir}\})} + (1-\pi_M){P_{RS}(\Yb_p|\text{par})},
\]
where 
\begin{itemize}
\item[]  in the first component responses are determined by model M with  $P_M(\Yb_p|\theta_p, \{\deltab_{ir}\})$ referring to a partial credit or some other ordinal  model,
\item[]  the second model $P_{RS}(\Yb_p|\text{par})$ specifies the response style that is suspected to be present.   
\end{itemize} 
For example, one might investigate if a portion of  respondents shows non-contingent  response style, which  is found if persons have a tendency to respond  carelessly, randomly, or nonpurposefully \citep{van2013response,baumgartner2001response} by specifying
\[
P_{RS}(\Yb_p|\text{par}= \prod_{i=1}^I  P(Y_{pi}|\{\deltab_{ir}^{m}\}),
\]
where $\{\deltab_{ir}^{m}\}$ are parameters that determine the marginal distribution of responses item $i$. The specification means that responses on items are independent and determined only by the item parameters.
An alternative is a mixture with the component
\[
P_{RS}(\Yb_p|\text{par}= P_M(\Yb_p|\theta_p, \{\deltab_{ir}^{RS}\}, \gamma_p),
\]
where $\gamma_p$ are additional response style parameters in a partial credit model if, for example, a partial credit model determines the first mixture component. The parameters $\{\deltab_{ir}^{RS}\}$ are parameters for model M in the second mixture component. Then it is assumed that respondents that are affected by response style have different parameters than respondents without response style. However, it may also be assume that the
parameters are the same as in the first component. Then one allows for respondents to be affected by response styles in differing degrees.  

Approaches like that go beyond the classical modelling of response styles in mixture models. In classical mixture model respondents are affected by response styles or not,  response style is considered  a discrete trait \citep{bolt2009addressing}. In contrast, in parametric models   response styles are represented by parameters, which may be small or large, varying across persons, and making response style  a continuous trait. The mixture given above combines these two worlds. Respondents may not be be affected by response style, or may be affected, but in different degrees.  

Models of this type seem not to have been considered, although there has been some development on modelling uncertainty, which is related to non-contingent 
response styles, however approaches were proposed mainly in the regression context not for repeated measurements as  item responses. For an overview of uncertainty modelling in regression see \citet{PIccSim}, repeated measures were considered by \citet{colombi2018hierarchical}.

\section{Concluding Remarks   }

It has been shown that an easily comprehensible taxonomy of ordinal item response models can be obtained by investigating the role of building blocks and split variables within the structure of ordinal models.
The  structure contains traditional models, IRTree models, and the   class of hierarchical partitioning models.
Although it is well known that ordinal models \textit{contain} binary models their role in the construction of ordinal models seems not to have been investigated in a systematic way to obtain a taxonomy. In particular  the distinction between non-conditional and conditional models,  the split of the latter into hierarchical and non-hierarchical models,   the role of the nominal model and how it is to be distinguished from ordinal models  contribute to  clarify the  structuring of polytomous models.

%Although it is   well known that IRTrees models are generalizations of sequential models the taxonomy embeds IRTrees into  a wider structure, and distinguishes between binary IRTrees and hierachical partitioning models.

%In particular the link to the split variables, which is important since it represents the use of order, has been used   to characterize adjacent categories models \citep{andrich2013expanded}, but not the whole class of ordinal models.

One of the advantages of having a distinct taxonomy of models is that  the meaning of parameters becomes clear. In particular,  effects in conditional models should be interpreted with regard to the conditioning, which holds for parametric and nonparametric approaches.
Parameterizations do not determine the taxonomy. They primarily determine  the complexity of the model, and specify which  effects are included in the model. 
Alternative parameterization have different meanings, and an additional slope parameter has a quite different interpretation if it is included in an adjacent categories model or an IRTree. Their meaning depends on the model type, and therefore on the placement in the taxonomy.

%Let us finally mention that there is an important class of models that has not been considered explicitly as part of the taxonomy, namely mixture models. They are mainly tools to investigate uni-dimensionality, and account for heterogeneity in responses and response styles. Originally developed for Rasch models by \citet{rost1991logistic}, they were extended to ordinal responses, among others, by \citet{rost1997applying}, \citet{eid2000detecting}, \citet{gollwitzer2005response}, \citet{maij2008fitting}, \citet{Moors2010},\citet{VanRos2010}, \citet{von2004partially}, for an overview see also \citet{von2007mixture}. Mixture models assume that the observed response results from a mixture of several components, the latent classes. Since in latent classes any polytomous model  can be used they can be seen as a sort of supermodel that comprises the class of polytomous models.

\begin{singlespace}
\bibliography{literatur}
\end{singlespace}
\end{document}

%% file: def.tex
\newcommand{\E}{\operatorname{E}}      %Erwartungswert
\newcommand{\var}{\operatorname{var}}
\newcommand{\cov}{\operatorname{cov}}

\newcommand{\deltab}{\boldsymbol{\delta}}

\newcommand{\Yb}{\boldsymbol{Y}}

\newcommand{\blanco}[1]{}

\def\d{\displaystyle}

%% file: TaxonomyPreprint.bbl
\begin{thebibliography}{}

\bibitem[\protect\citeauthoryear{Adams, Wu, and Wilson}{Adams
  et~al.}{2012}]{adams2012rasch}
Adams, R.~J., M.~L. Wu, and M.~Wilson (2012).
\newblock The {R}asch rating model and the disordered threshold controversy.
\newblock {\em Educational and Psychological Measurement\/}~{\em 72\/}(4),
  547--573.

\bibitem[\protect\citeauthoryear{Agresti}{Agresti}{2013}]{Agresti:2013}
Agresti, A. (2013).
\newblock {\em Categorical Data Analysis, 3d Edition}.
\newblock New York: Wiley.

\bibitem[\protect\citeauthoryear{Andersen}{Andersen}{1977}]{Andersen:77}
Andersen, E.~B. (1977).
\newblock Sufficient statistics and latent trait models.
\newblock {\em Psychometrika\/}~{\em 42}, 69--81.

\bibitem[\protect\citeauthoryear{Andrich}{Andrich}{1978}]{andrich1978rating}
Andrich, D. (1978).
\newblock A rating formulation for ordered response categories.
\newblock {\em Psychometrika\/}~{\em 43\/}(4), 561--573.

\bibitem[\protect\citeauthoryear{Andrich}{Andrich}{2010}]{andrich2010sufficiency}
Andrich, D. (2010).
\newblock Sufficiency and conditional estimation of person parameters in the
  polytomous {R}asch model.
\newblock {\em Psychometrika\/}~{\em 75\/}(2), 292--308.

\bibitem[\protect\citeauthoryear{Andrich}{Andrich}{2013}]{andrich2013expanded}
Andrich, D. (2013).
\newblock An expanded derivation of the threshold structure of the polytomous
  {R}asch model that dispels any 'threshold disorder controversy'.
\newblock {\em Educational and Psychological Measurement\/}~{\em 73\/}(1),
  78--124.

\bibitem[\protect\citeauthoryear{Andrich}{Andrich}{2015}]{andrich2015problem}
Andrich, D. (2015).
\newblock The problem with the step metaphor for polytomous models for ordinal
  assessments.
\newblock {\em Educational Measurement: Issues and Practice\/}~{\em 34\/}(2),
  8--14.

\bibitem[\protect\citeauthoryear{Andrich}{Andrich}{2016}]{Andrichh2016}
Andrich, D. (2016).
\newblock {R}asch rating-scale model.
\newblock In W.~Van~der Linden (Ed.), {\em Handbook of {M}odern {I}tem
  {R}esponse {T}heory}, pp.\  75--94. Springer.

\bibitem[\protect\citeauthoryear{Baumgartner and Steenkamp}{Baumgartner and
  Steenkamp}{2001}]{baumgartner2001response}
Baumgartner, H. and J.-B.~E. Steenkamp (2001).
\newblock Response styles in marketing research: A cross-national
  investigation.
\newblock {\em Journal of Marketing Research\/}~{\em 38\/}(2), 143--156.

\bibitem[\protect\citeauthoryear{Bock}{Bock}{1972}]{bock1972estimating}
Bock, R.~D. (1972).
\newblock Estimating item parameters and latent ability when responses are
  scored in two or more nominal categories.
\newblock {\em Psychometrika\/}~{\em 37\/}(1), 29--51.

\bibitem[\protect\citeauthoryear{B{\"o}ckenholt}{B{\"o}ckenholt}{2012}]{bockenholt2013modeling}
B{\"o}ckenholt, U. (2012).
\newblock Modeling multiple response processes in judgment and choice.
\newblock {\em Psychological Methods\/}~{\em 17\/}(4), 665--678.

\bibitem[\protect\citeauthoryear{B{\"o}ckenholt}{B{\"o}ckenholt}{2017}]{bockenholt2016measuring}
B{\"o}ckenholt, U. (2017).
\newblock Measuring response styles in {L}ikert items.
\newblock {\em Psychological methods\/}~(22), 69--83.

\bibitem[\protect\citeauthoryear{B{\"o}ckenholt and Meiser}{B{\"o}ckenholt and
  Meiser}{2017}]{bockenholt2017response}
B{\"o}ckenholt, U. and T.~Meiser (2017).
\newblock Response style analysis with threshold and multi-process irt models:
  A review and tutorial.
\newblock {\em British Journal of Mathematical and Statistical
  Psychology\/}~{\em 70\/}(1), 159--181.

\bibitem[\protect\citeauthoryear{Bolt and Johnson}{Bolt and
  Johnson}{2009}]{bolt2009addressing}
Bolt, D.~M. and T.~R. Johnson (2009).
\newblock Addressing score bias and differential item functioning due to
  individual differences in response style.
\newblock {\em Applied Psychological Measurement\/}~{\em 33\/}(5), 335--352.

\bibitem[\protect\citeauthoryear{Colombi, Giordano, Gottard, and
  Iannario}{Colombi et~al.}{2018}]{colombi2018hierarchical}
Colombi, R., S.~Giordano, A.~Gottard, and M.~Iannario (2018).
\newblock Hierarchical marginal models with latent uncertainty.
\newblock {\em Scandinavian Journal of Statistics, to appear\/}.

\bibitem[\protect\citeauthoryear{De~Boeck, Cho, and Wilson}{De~Boeck
  et~al.}{2011}]{de2011explanatory}
De~Boeck, P., S.-J. Cho, and M.~Wilson (2011).
\newblock Explanatory secondary dimension modeling of latent differential item
  functioning.
\newblock {\em Applied Psychological Measurement\/}~{\em 35\/}(8), 583--603.

\bibitem[\protect\citeauthoryear{De~Boeck and Partchev}{De~Boeck and
  Partchev}{2012}]{de2012irtrees}
De~Boeck, P. and I.~Partchev (2012).
\newblock Irtrees: Tree-based item response models of the glmm family.
\newblock {\em Journal of Statistical Software\/}~{\em 48\/}(1), 1--28.

\bibitem[\protect\citeauthoryear{Eid and Rauber}{Eid and
  Rauber}{2000}]{eid2000detecting}
Eid, M. and M.~Rauber (2000).
\newblock Detecting measurement invariance in organizational surveys.
\newblock {\em European Journal of Psychological Assessment\/}~{\em 16\/}(1),
  20--30.

\bibitem[\protect\citeauthoryear{Garc{\'\i}a-P{\'e}rez}{Garc{\'\i}a-P{\'e}rez}{2017}]{garcia2017analysis}
Garc{\'\i}a-P{\'e}rez, M.~A. (2017).
\newblock An analysis of (dis) ordered categories, thresholds, and crossings in
  difference and divide-by-total irt models for ordered responses.
\newblock {\em The Spanish Journal of Psychology\/}~{\em 20}, 1--27.

\bibitem[\protect\citeauthoryear{Gollwitzer, Eid, and J{\"u}rgensen}{Gollwitzer
  et~al.}{2005}]{gollwitzer2005response}
Gollwitzer, M., M.~Eid, and R.~J{\"u}rgensen (2005).
\newblock Response styles in the assessment of anger expression.
\newblock {\em Psychological assessment\/}~{\em 17\/}(1), 56.

\bibitem[\protect\citeauthoryear{Hemker, Sijtsma, Molenaar, and Junker}{Hemker
  et~al.}{1997}]{hemker1997stochastic}
Hemker, B.~T., K.~Sijtsma, I.~W. Molenaar, and B.~W. Junker (1997).
\newblock Stochastic ordering using the latent trait and the sum score in
  polytomous {IRT} models.
\newblock {\em Psychometrika\/}~{\em 62\/}(3), 331--347.

\bibitem[\protect\citeauthoryear{Hemker, van~der Ark, and Sijtsma}{Hemker
  et~al.}{2001}]{hemker2001measurement}
Hemker, B.~T., L.~A. van~der Ark, and K.~Sijtsma (2001).
\newblock On measurement properties of continuation ratio models.
\newblock {\em Psychometrika\/}~{\em 66\/}(4), 487--506.

\bibitem[\protect\citeauthoryear{Jansen and Roskam}{Jansen and
  Roskam}{1986}]{jansen1986latent}
Jansen, P.~G. and E.~E. Roskam (1986).
\newblock Latent trait models and dichotomization of graded responses.
\newblock {\em Psychometrika\/}~{\em 51\/}(1), 69--91.

\bibitem[\protect\citeauthoryear{Jeon and De~Boeck}{Jeon and
  De~Boeck}{2016}]{jeon2016generalized}
Jeon, M. and P.~De~Boeck (2016).
\newblock A generalized item response tree model for psychological assessments.
\newblock {\em Behavior research methods\/}~{\em 48\/}(3), 1070--1085.

\bibitem[\protect\citeauthoryear{Jin and Wang}{Jin and
  Wang}{2014}]{jin2014generalized}
Jin, K.-Y. and W.-C. Wang (2014).
\newblock Generalized irt models for extreme response style.
\newblock {\em Educational and Psychological Measurement\/}~{\em 74\/}(1),
  116--138.

\bibitem[\protect\citeauthoryear{Johnson}{Johnson}{2003}]{johnson2003use}
Johnson, T.~R. (2003).
\newblock On the use of heterogeneous thresholds ordinal regression models to
  account for individual differences in response style.
\newblock {\em Psychometrika\/}~{\em 68\/}(4), 563--583.

\bibitem[\protect\citeauthoryear{Khorramdel and von Davier}{Khorramdel and von
  Davier}{2014}]{khorramdel2014measuring}
Khorramdel, L. and M.~von Davier (2014).
\newblock Measuring response styles across the big five: A multiscale extension
  of an approach using multinomial processing trees.
\newblock {\em Multivariate Behavioral Research\/}~{\em 49\/}(2), 161--177.

\bibitem[\protect\citeauthoryear{Kulas, Stachowski, and Haynes}{Kulas
  et~al.}{2008}]{kulas2008middle}
Kulas, J.~T., A.~A. Stachowski, and B.~A. Haynes (2008).
\newblock Middle response functioning in likert-responses to personality items.
\newblock {\em Journal of Business and Psychology\/}~{\em 22\/}(3), 251--259.

\bibitem[\protect\citeauthoryear{Maij-de Meij, Kelderman, and van~der
  Flier}{Maij-de Meij et~al.}{2008}]{maij2008fitting}
Maij-de Meij, A.~M., H.~Kelderman, and H.~van~der Flier (2008).
\newblock Fitting a mixture item response theory model to personality
  questionnaire data: Characterizing latent classes and investigating
  possibilities for improving prediction.
\newblock {\em Applied Psychological Measurement\/}~{\em 32\/}(8), 611--631.

\bibitem[\protect\citeauthoryear{Masters}{Masters}{1982}]{Masters:82}
Masters, G.~N. (1982).
\newblock A {R}asch model for partial credit scoring.
\newblock {\em Psychometrika\/}~{\em 47}, 149--174.

\bibitem[\protect\citeauthoryear{Masters and Wright}{Masters and
  Wright}{1984}]{MasWri:84}
Masters, G.~N. and B.~Wright (1984).
\newblock The essential process in a family of measurement models.
\newblock {\em Psychometrika\/}~{\em 49}, 529--544.

\bibitem[\protect\citeauthoryear{Meiser, Plieninger, and Henninger}{Meiser
  et~al.}{2019}]{meiser2019irt}
Meiser, T., H.~Plieninger, and M.~Henninger (2019).
\newblock Irt ree models with ordinal and multidimensional decision nodes for
  response styles and trait-based rating responses.
\newblock {\em British Journal of Mathematical and Statistical Psychology\/}.

\bibitem[\protect\citeauthoryear{Mokken}{Mokken}{1971}]{mokken1971theory}
Mokken, R.~J. (1971).
\newblock {\em A theory and procedure of scale analysis}.
\newblock Berlin: Walter de Gruyter.

\bibitem[\protect\citeauthoryear{Moors}{Moors}{2010}]{Moors2010}
Moors, G. (2010).
\newblock Ranking the ratings: A latent-class regression model to control for
  overall agreement in opinion research.
\newblock {\em International Journal of Public Opinion Research\/}~{\em
  22\/}(1), 93--119.

\bibitem[\protect\citeauthoryear{Muraki}{Muraki}{1990}]{muraki1990fitting}
Muraki, E. (1990).
\newblock Fitting a polytomous item response model to {L}ikert-type data.
\newblock {\em Applied Psychological Measurement\/}~{\em 14\/}(1), 59--71.

\bibitem[\protect\citeauthoryear{Muraki}{Muraki}{1997}]{muraki1997generalized}
Muraki, E. (1997).
\newblock A generalized partial credit model.
\newblock {\em Handbook of modern item response theory\/}, 153--164.

\bibitem[\protect\citeauthoryear{Piccolo and Simone}{Piccolo and
  Simone}{2019}]{PIccSim}
Piccolo, D. and R.~Simone (2019).
\newblock The class of {CUB} models: statistical foundations, inferential
  issues and empirical evidence.
\newblock {\em Statistical Methods and Applications,
  https://doi.org/10.1007/s10260-019-00461-1\/}.

\bibitem[\protect\citeauthoryear{Plieninger}{Plieninger}{2016}]{plieninger2016mountain}
Plieninger, H. (2016).
\newblock Mountain or molehill? a simulation study on the impact of response
  styles.
\newblock {\em Educational and Psychological Measurement\/}~{\em 77}, 32--53.

\bibitem[\protect\citeauthoryear{Plieninger}{Plieninger}{2020}]{plieninger2020developing}
Plieninger, H. (2020).
\newblock Developing and applying ir-tree models: Guidelines, caveats, and an
  extension to multiple groups.
\newblock {\em Organizational Research Methods,
  doi:10.1177/1094428120911096\/}.

\bibitem[\protect\citeauthoryear{Rost}{Rost}{1991}]{rost1991logistic}
Rost, J. (1991).
\newblock A logistic mixture distribution model for polychotomous item
  responses.
\newblock {\em British Journal of Mathematical and Statistical
  Psychology\/}~{\em 44\/}(1), 75--92.

\bibitem[\protect\citeauthoryear{Rost, Carstensen, and Von~Davier}{Rost
  et~al.}{1997}]{rost1997applying}
Rost, J., C.~Carstensen, and M.~Von~Davier (1997).
\newblock Applying the mixed rasch model to personality questionnaires.
\newblock {\em Applications of latent trait and latent class models in the
  social sciences\/}, 324--332.

\bibitem[\protect\citeauthoryear{Samejima}{Samejima}{1995}]{samejima1995acceleration}
Samejima, F. (1995).
\newblock Acceleration model in the heterogeneous case of the general graded
  response model.
\newblock {\em Psychometrika\/}~{\em 60\/}(4), 549--572.

\bibitem[\protect\citeauthoryear{Samejima}{Samejima}{2016}]{samejima2016graded}
Samejima, F. (2016).
\newblock Graded response model.
\newblock In W.~Van~der Linden (Ed.), {\em Handbook of item response theory},
  pp.\  95--108.

\bibitem[\protect\citeauthoryear{Shu, Henson, and Luecht}{Shu
  et~al.}{2013}]{shu2013using}
Shu, Z., R.~Henson, and R.~Luecht (2013).
\newblock Using deterministic, gated item response theory model to detect test
  cheating due to item compromise.
\newblock {\em Psychometrika\/}~{\em 78\/}(3), 481--497.

\bibitem[\protect\citeauthoryear{Sijtsma and Hemker}{Sijtsma and
  Hemker}{2000}]{sijtsma2000taxonomy}
Sijtsma, K. and B.~T. Hemker (2000).
\newblock A taxonomy of irt models for ordering persons and items using simple
  sum scores.
\newblock {\em Journal of Educational and Behavioral Statistics\/}~{\em
  25\/}(4), 391--415.

\bibitem[\protect\citeauthoryear{Sijtsma and Molenaar}{Sijtsma and
  Molenaar}{2002}]{sijtsma2002introduction}
Sijtsma, K. and I.~W. Molenaar (2002).
\newblock {\em Introduction to nonparametric item response theory}, Volume~5.
\newblock Sage.

\bibitem[\protect\citeauthoryear{Thissen and Cai}{Thissen and
  Cai}{2016}]{ThissCai2016}
Thissen, D. and L.~Cai (2016).
\newblock Nominal categories model.
\newblock In W.~Van~der Linden (Ed.), {\em Handbook of {M}odern {I}tem
  {R}esponse {T}heory}, pp.\  51--73. Springer.

\bibitem[\protect\citeauthoryear{Thissen, Cai, and Bock}{Thissen
  et~al.}{2010}]{thissen2010nominal}
Thissen, D., L.~Cai, and R.~D. Bock (2010).
\newblock The nominal categories item response model.
\newblock {\em Handbook of polytomous item response theory models\/}, 43--75.

\bibitem[\protect\citeauthoryear{Thissen and Steinberg}{Thissen and
  Steinberg}{1986}]{thissen1986taxonomy}
Thissen, D. and L.~Steinberg (1986).
\newblock A taxonomy of item response models.
\newblock {\em Psychometrika\/}~{\em 51\/}(4), 567--577.

\bibitem[\protect\citeauthoryear{Thissen-Roe and Thissen}{Thissen-Roe and
  Thissen}{2013}]{thissen2013two}
Thissen-Roe, A. and D.~Thissen (2013).
\newblock A two-decision model for responses to {L}ikert-type items.
\newblock {\em Journal of Educational and Behavioral Statistics\/}~{\em
  38\/}(5), 522--547.

\bibitem[\protect\citeauthoryear{Tijmstra, Bolsinova, and Jeon}{Tijmstra
  et~al.}{2018}]{tijmstra2018generalized}
Tijmstra, J., M.~Bolsinova, and M.~Jeon (2018).
\newblock Generalized mixture irt models with different item-response
  structures: A case study using {L}ikert-scale data.
\newblock {\em Behavior Research Methods\/}~{\em 55}, 1--20.

\bibitem[\protect\citeauthoryear{Tutz}{Tutz}{1990}]{Tutz:90b}
Tutz, G. (1990).
\newblock Sequential item response models with an ordered response.
\newblock {\em British Journal of Statistical and Mathematical
  Psychology\/}~{\em 43}, 39--55.

\bibitem[\protect\citeauthoryear{Tutz}{Tutz}{2020a}]{TuLikert2020}
Tutz, G. (2020a).
\newblock Hierarchical models for the analysis of {L}ikert scales in regression
  and item response analysis.
\newblock {\em International Statistical Review, doi:10.1111/insr.12396\/}.

\bibitem[\protect\citeauthoryear{Tutz}{Tutz}{2020b}]{Tu2020JMP}
Tutz, G. (2020b).
\newblock On the structure of ordered latent trait models.
\newblock {\em Journal of Mathematical Psychology\/}~{\em 96}.

\bibitem[\protect\citeauthoryear{Tutz and Draxler}{Tutz and
  Draxler}{2019}]{TuDrax2019}
Tutz, G. and C.~Draxler (2019).
\newblock A common framework for classical and tree-based item response models
  including extended hierarchically structured models.
\newblock Technical Report 227, Department of Statistics LMU Munich.

\bibitem[\protect\citeauthoryear{Tutz, Schauberger, and Berger}{Tutz
  et~al.}{2018}]{TuSchBe2018}
Tutz, G., G.~Schauberger, and M.~Berger (2018).
\newblock Response styles in the partial credit model.
\newblock {\em Applied Psychological Measurement\/}~{\em 42}, 407--427.

\bibitem[\protect\citeauthoryear{Van~der Linden}{Van~der
  Linden}{2016}]{VanderLind2016}
Van~der Linden, W. (2016).
\newblock {\em Handbook of {I}tem {R}esponse {T}heory}.
\newblock Springer: New York.

\bibitem[\protect\citeauthoryear{Van~Rosmalen, Van~Herk, and
  Groenen}{Van~Rosmalen et~al.}{2010}]{VanRos2010}
Van~Rosmalen, J., H.~Van~Herk, and P.~Groenen (2010).
\newblock Identifying response styles: A latent-class bilinear multinomial
  logit model.
\newblock {\em Journal of Marketing Research\/}~{\em 47\/}(1), 157--172.

\bibitem[\protect\citeauthoryear{Van~Vaerenbergh and Thomas}{Van~Vaerenbergh
  and Thomas}{2013}]{van2013response}
Van~Vaerenbergh, Y. and T.~D. Thomas (2013).
\newblock Response styles in survey research: A literature review of
  antecedents, consequences, and remedies.
\newblock {\em International Journal of Public Opinion Research\/}~{\em
  25\/}(2), 195--217.

\bibitem[\protect\citeauthoryear{Verhelst, Glas, and De~Vries}{Verhelst
  et~al.}{1997}]{verhelst1997steps}
Verhelst, N.~D., C.~Glas, and H.~De~Vries (1997).
\newblock A steps model to analyze partial credit.
\newblock In {\em Handbook of modern item response theory}, pp.\  123--138.
  Springer.

\bibitem[\protect\citeauthoryear{Von~Davier}{Von~Davier}{1996}]{von1996mixtures}
Von~Davier, M. (1996).
\newblock Mixtures of polytomous rasch models and latent class models for
  ordinal variables.
\newblock {\em Softstat\/}~{\em 95}.

\bibitem[\protect\citeauthoryear{Von~Davier and Yamamoto}{Von~Davier and
  Yamamoto}{2004}]{von2004partially}
Von~Davier, M. and K.~Yamamoto (2004).
\newblock Partially observed mixtures of {IRT} models: An extension of the
  generalized partial-credit model.
\newblock {\em Applied Psychological Measurement\/}~{\em 28\/}(6), 389--406.

\bibitem[\protect\citeauthoryear{Von~Davier and Yamamoto}{Von~Davier and
  Yamamoto}{2007}]{von2007mixture}
Von~Davier, M. and K.~Yamamoto (2007).
\newblock Mixture-distribution and hybrid rasch models.
\newblock In {\em Multivariate and mixture distribution Rasch models}, pp.\
  99--115. Springer.

\bibitem[\protect\citeauthoryear{Wetzel and Carstensen}{Wetzel and
  Carstensen}{2017}]{wetzel2015multidimensional}
Wetzel, E. and C.~H. Carstensen (2017).
\newblock Multidimensional modeling of traits and response styles.
\newblock {\em European Journal of Psychological Assessment\/}~(33), 352--364.

\end{thebibliography}
